\documentclass{aa}

\usepackage{graphicx}
\usepackage{txfonts}
\begin{document}

   \title{EeV Astrophysical neutrinos from FSRQs?}

   \author{C. Righi
          \inst{1} 
          \and
          A. Palladino\inst{2}
          \and
          F. Tavecchio
          \inst{1}
          \and
          F. Vissani
          \inst{3} \fnmsep \inst{4}
          }

   \institute{INAF -- Osservatorio Astronomico di Brera, via E. Bianchi 46, I--23807 Merate, Italy
         \and
             DESY, Platanenallee 6, 15738 Zeuthen, Germany 
            \and
              Gran Sasso Science Institute (GSSI), Viale F. Crispi 7, 67100 L’Aquila, Italy
              \and
INFN, Laboratori Nazionali del Gran Sasso (LNGS), 67100 Assergi, L’Aquila, Italy\\
\email{chiara.righi@inaf.it}
}


\abstract
   {Flat Spectrum Radio Quasars (FSRQ) are the most powerful blazars in the $\gamma$-ray band. Although they are supposed to be good candidates in producing high energy neutrinos, no secure detection of FSRQs has been obtained up to now, except for a possible case of PKS B1424-418.}
   {In this work we compute the expected flux of high energy neutrinos from FSRQs using standard assumptions for the properties of the radiation fields filling the regions surrounding the central supermassive black hole.}
   {Starting from the FSRQ spectral sequence, we compute the neutrino spectrum assuming interaction of relativistic protons with internal and external radiation fields. We study the  neutrino spectra resulting from different values of free parameters.}
   {We obtain as a result that high energy neutrinos are naturally expected from FSRQs in the sub-EeV--EeV energy range and not at PeV energies. This justifies the non-observation of neutrinos from FSRQs with the present technology, since only neutrinos below 10 PeV have been observed. We found that for a non-negligible range of the parameters the cumulative flux from FSRQs is comparable to or even exceeds the expected cosmogenic neutrino flux. This result is intriguing and highlights the importance to disentangle these point-source emission from the diffuse cosmogenic background.}
   {}

\maketitle


\section{Introduction}

The extreme phenomenology displayed by Active galactic nuclei (AGN) is ultimately associated with the gravitational energy released by gas falling onto a supermassive black hole ($M_{\rm BH}=10^{7-10} M_{\odot}$) residing at the center of the host galaxies. AGN with relativistic jets, although a minor fraction of the total population, represent the most powerful persistent sources of electromagnetic radiation in the Universe \citep{BlandfordreviewAGN}. Blazars \citep{Romero17} are a subclass of jetted AGNs in which the jet is closely aligned with the line of sight. Under this favorable geometry, the non-thermal emission of the jet is highly amplified by relativistic effects (relativistic beaming). The observed electromagnetic emission of these sources extends over the entire electromagnetic spectrum and it is characterized by strong variability that can enhance the luminosity by several orders of magnitude during flares. The spectral energy distribution (SED) of blazars -- dominated by the non-thermal emission of the relativistic jet --  typically shows a double bump shape. The first peak is in the IR-UV band, the high-energy one has the maximum in the $\gamma$-ray band. While the first bump originates from the synchrotron emission of relativistic electron in the jet, the physical interpretation of the second peak is still under debate, also in view of the neutrino detection associated with one of these objects \citep{IceCube:2018dnn}. In the pure leptonic model gamma rays are produced through the inverse Compton scattering by the synchrotron-emitting relativistic electrons. In pure hadronic scenarios, on the other hand, high-energy photons are supposed to be produced through synchrotron emission of relativistic protons \citep{Aharonian00} or the reprocessed radiation produced in photomeson interactions (e.g. \citealt{Aharonian00}, \citealt{Mucke2003}).

Blazars can be divided into two subclasses: Flat Spectrum Radio Quasars (FSRQ) and BL Lac objects. Their classification is historically based on the width of the emission lines observed in the optical spectra (the traditional division corresponds to an equivalent width of 5 \AA). FSRQs display broad lines and are generally more powerful than BL Lac, especially in the $\gamma$-ray band. On the other hand, BL Lacs, although less powerful, can emit photons at higher energies (up to the multi-TeV band). At present, there are pieces of evidence supporting the existence of a continuous trend (or sequence) between the SED of the two subclasses (see \citealt{Fossati98}, \citealt{GG17}, G17 hereafter).

The detection by IceCube of a high energy neutrino ($E_{\nu}\sim 290$ TeV) in the direction of a flaring blazar (whose precise nature is debated, see \citealt{Padovani:2019xcv}), TXS 0506+056, on September 2017 \citep{IceCube:2018dnn}, has lead the community to focus on this class as electromagnetic counterparts of astrophysical neutrinos (\citealt{Palladino:2018lov, Murase:2018iyl,Padovani:2016wwn}), although many other candidates have been proposed (see \citealt{Ahalers14}, \citealt{Meszaros17} or \citealt{Gaisser18} for reviews). The production of neutrinos requires the existence of a population of relativistic protons (with energies $E_{\rm p}\simeq 20 \times E_{\nu}$) creating charge pions in collisions with target protons/ions or photons (the latter case is usually referred as photomeson reaction). Relativistic protons are thought to be co-accelerated with the electrons, although the details of the mechanisms involved are not completely clear. Considering the low particle density inferred for the relativistic outflows, pion production through the collision with gas is widely considered unlikely (but see e.g. \citealt{Sahakyan18} for possible scenarios) and, therefore, neutrino production is expected to proceed mainly through the photomeson channel. This reaction is characterized by a well-defined threshold for energies very close to that corresponding to the $\Delta$ resonance, ($p\gamma\to \Delta^{+}$ followed by the quick decay to $p\pi^0$ or $n\pi^{+}$) and this allows one to derive a useful rule-of-thumb relating the energy of the target photons involved in the reaction, $\epsilon$, and the proton energy, $E_{\rm p}$, as: $E_{\rm p}\simeq 1.3\times 10^{17}/\epsilon$ eV (where $\epsilon$ is expressed in eV). From this relation, and taking into account that the typical neutrino energy is $1/20$ of that of the parent proton, one can link the efficient production of neutrinos with energy in the range 0.1-10 PeV with target photons in the optical-soft X-ray bands. 

While suggested to be potential good neutrino emitters (e.g. \citealt{Murase12}, \citealt{Dermer:2014vaa,Kadler:2016ygj}), FSRQ are not favoured by current data. In particular, being powerful but relatively rare in the sky, one should expect to detect multiplets of neutrino events from single FSRQ already in the current samples (e.g. \citealt{MuraseWaxman16}). This rules out the possibility that FSRQs represent the bulk of the sources of the neutrinos detected by IceCube, in the 0.1-10 PeV energy range. Interestingly, this conclusion is consistent with the current picture of the structure of FSRQ. In fact, the most intense radiation field expected to interact with relativistic protons in the inner regions of the jet has typical frequencies corresponding to the UV-optical-IR bands, corresponding to relatively large $E_{\rm p}$ which, in turn, imply $E_{\nu}$ above the PeV band.
In view of the physics programs of existing and future neutrino telescopes, this expectation deserves to be discussed carefully and quantitatively.

In this work we focus on FSRQ using the state-of-the-art knowledge on their structure and demography to infer their potential neutrino emission. In section \ref{sec:FSRQ} we describe the population and the structure of FSRQs we consider in the work, section \ref{sec:neutrinos} describes the calculation of the neutrino emission we used, while section \ref{sec:models} describes the set-up of our models.
We report the resulting cumulative predicted spectra in \ref{sec:results} in which we also compared our results with the most updated IceCube data, the expected spectra for cosmogenic neutrinos and the sensibility of the future telescope GRAND. Finally in section \ref{sec:discussion} we conclude with a discussion.

\section{Properties of FSRQs}
\label{sec:FSRQ}
\begin{figure}
\centering
\includegraphics[width=0.45\textwidth]{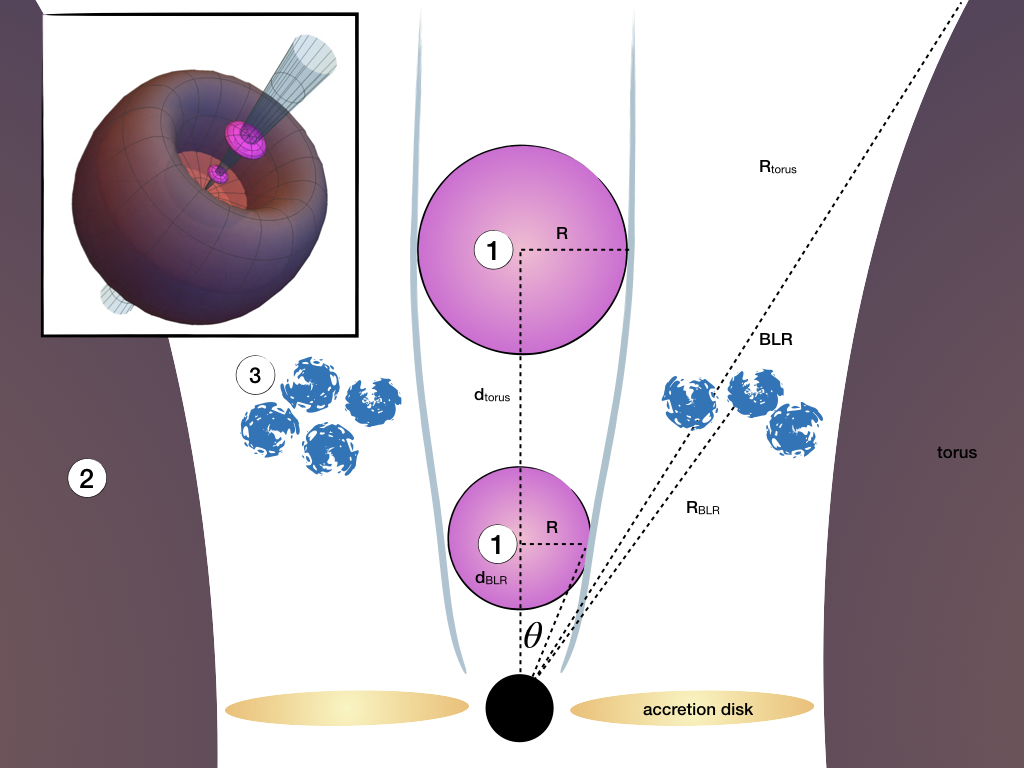}
\caption{Sketch of the geometry we assumed for FSRQ. Region 1 is the emission region containing cosmic rays and photons from the synctrotron emission of electrons. Region 2 and region 3 are the torus and the broad line region that produce external photon populations. See text for a more detailed description.}
\label{fig:sketch_FSRQ}
\end{figure}

\subsection{Structure}

A sketch of the structure of the central regions (distances less then few pc) of FSRQ is reported in Fig.\ref{fig:sketch_FSRQ}. We represent the different region of interest in our work. 
Accretion of gas occurs through a standard, optically thick and geometrically thin accretion disk, showed in yellow in fig.\ref{fig:sketch_FSRQ} \citep{Shakura97}. The emission from the disk, peaking in the optical-UV band, irradiates small clouds of gas moving in keplerian orbits around the SMBH. The photoionized gas of the clouds re-emit the intercepted flux in prominent emission lines (with average energy $\epsilon_{\rm BLR}\simeq 10$ eV), whose observed profile is broadened by Doppler shift. The (usually assumed shell-like) region filled by these clouds is generally named Broad Line Region, BLR, represented in blue in the sketch of figure \ref{fig:sketch_FSRQ}. The geometry of the BLR can be probed by the reverberation mapping technique, which allows direct measurements of the distance of the clouds \citep{Kaspi07}. Beyond the BLR, at a typical distances 1-10 pc we find a thick structure of dust, traditionally assumed to have a toroidal shape (region 2 of figure \ref{fig:sketch_FSRQ} in purple). Dust grains are powerful absorbers of radiation and the torus obscures the central regions for observers lying at large angles. Dust heated by the central UV radiation re-emit it as a black-body spectrum at a temperature $T_{\rm IR}\simeq 500-10^3$ K, peaking in the IR band. 

As discussed in \cite{GGTF09} an approximated estimate of the typical size of the BLR ($R_{\rm BLR}$) and the torus ($R_{\rm torus}$) can be expressed as a function of the disk luminosity $L_{\rm d}$ as:
\begin{equation}
    R_{\rm BLR}\simeq 10^{17} \left( \frac{L_{\rm d}}{10^{45} \; {\rm erg/s}} \right)^{1/2} \;{\rm cm} \\
\label{eq:rblr}
\end{equation}
and
\begin{equation}
    R_{\rm torus}\simeq 2.5\times 10^{18} \left( \frac{L_{\rm d}}{10^{45} \; {\rm erg/s}} \right)^{1/2} \;{\rm cm}.
    \label{eq:rtorus}
\end{equation}
These distances are very important since, together with the corresponding luminosities, determine the photon density, a critical parameter for the efficiency of the photopion reactions. Note that, since the radiation energy density is proportional to $L/R^2$, the simple scaling above implies {\it fixed energy densities for all $L_{\rm d}$, i.e. for all FSRQ}.  

The jet is launched in the vicinity of the SMBH and it is progressively accelerated and collimated. The physical processes at the base of jet formation and acceleration are still under debate (see for a review \citealt{BlandfordreviewAGN} or \citealt{Romero}). The scenario outlined by MHD numerical simulations (e.g. \citealt{McKinney06}) describes a jet that starts as a magnetically dominated (Poynting) flow that accelerate progressively converting the magnetic energy flux into kinetic energy flux. The acceleration is relatively low: the flow is expected to reach the final Lorentz factor (when magnetic and kinetic powers are almost in equipartition) around the parsec scale (e.g. \citealt{Komissarov07}, \citealt{Vlahakis15}, \citealt{Nakamura}). This is the region generally identified with the ``blazar region", i.e. the region where the bulk of the radiation observed from blazar is produced. 

Jets of FSRQ propagate through the different structures discussed above. In the reference frame of the outflowing plasma, moving with constant bulk Lorentz factor $\Gamma$, the energy density of the radiation fields is modified by the relativistic boosting. At distances for which the jet is totally immersed in the (nearly isotropic) radiation field of the BLR or of the torus, the relativistic boosting results in the shift of the frequencies by a factor $\approx \Gamma$ and the amplification of the energy density by a factor $\approx \Gamma^2$ with respect to the value measured by an external observer at rest (e.g. \citealt{Sikora94}). Although the spectrum emitted by the BLR clouds is rather complex, with the presence of lines and continuum, the boosted emission as observed in the jet rest frame can well approximated by a peaked, black-body like spectrum with the maximum corresponding to the Lyman $\alpha$ line, i.e. at an energy $\epsilon^{\prime}_{\rm BLR}\approx 10 \, \Gamma$ eV \citep{FTGG08}. In the calculation of the neutrino spectra we will use this approximation.

\subsection{Sites of neutrino production}

The radiation we observe from FSRQ emerges from an emission region inside the jet. For simplicity we assume a  spherical region, at a distance $d$ from the central SMBH and with a radius $R$ as reported in figure \ref{fig:sketch_FSRQ}. We assume a conical jet with semi-aperture angle $\theta_{\rm j}$, so that $R=d\theta_{\rm j}$. The location of the emission region is FSRQ is a subject of debate. While standard models assume that the emission mainly occurs at distances lower than $R_{\rm BLR}$ (thus within the dense radiation field of the BLR), the absence of spectral features related to absorption of gamma rays by the UV continuum (e.g. \citealt{Poutanen10}, \citealt{Costamante18}), together with the detection of few FSRQ at TeV energies during flares (e.g. \citealt{MAGIC_PKS1510-089} or \citealt{HESS_3C279}) indicates that, at least in some cases and/or during high-activity states, the electromagnetic radiation is produced beyond the BLR (e.g. \citealt{FT11}). 

In the following we will calculate the neutrino output expected for two cases, namely assuming the emission region located at a distance $d<R_{\rm BLR}$ and $R_{\rm BLR}<d<R_{\rm torus}$. We will distinguish the two scenarios calling them ``inside BLR'' and ``outside BLR''. For the torus we assume $T_{\rm IR}=500$ K.

\begin{figure*}
\centering
\includegraphics[width=0.30\textwidth]{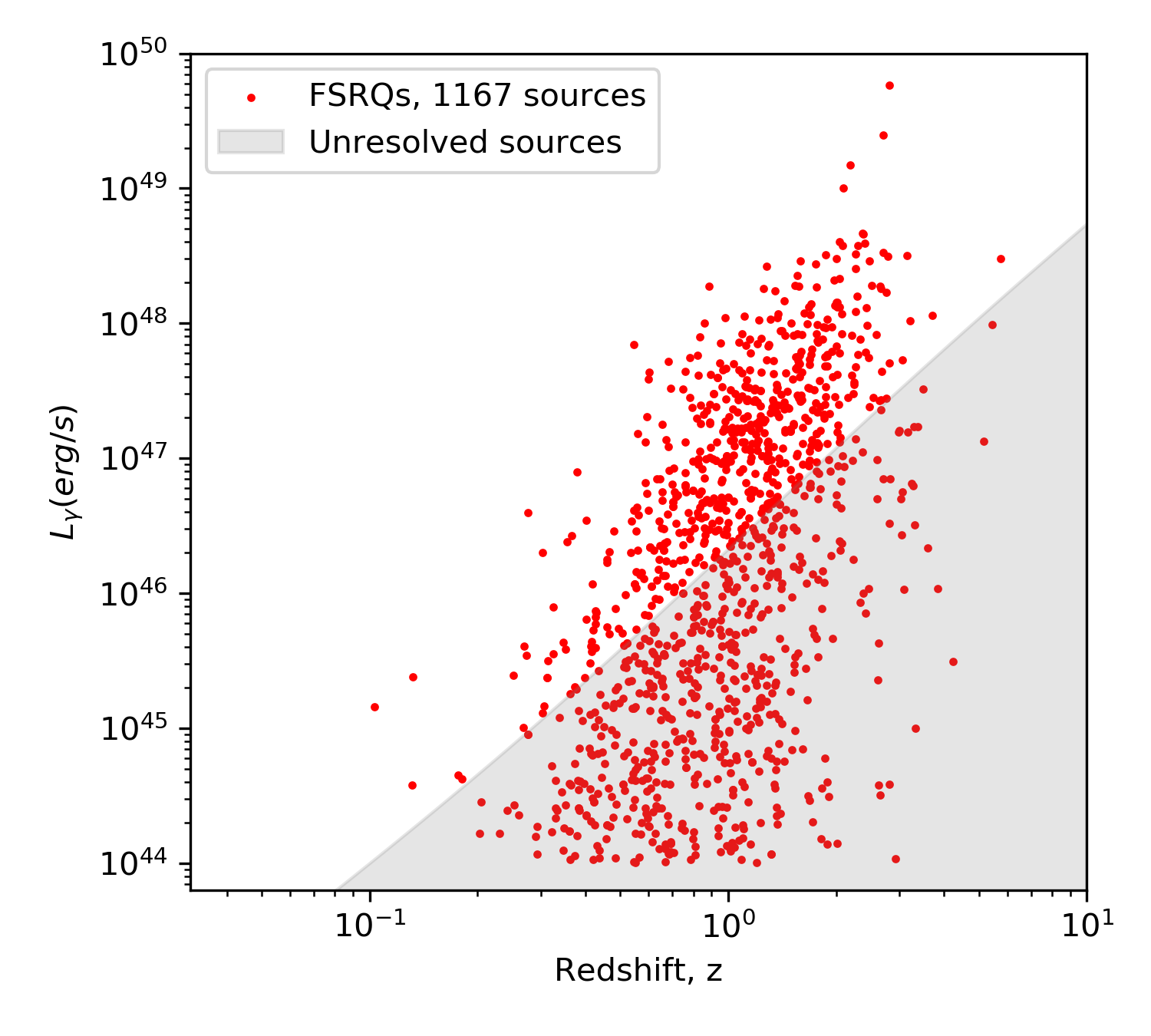}
\includegraphics[width=0.33\textwidth]{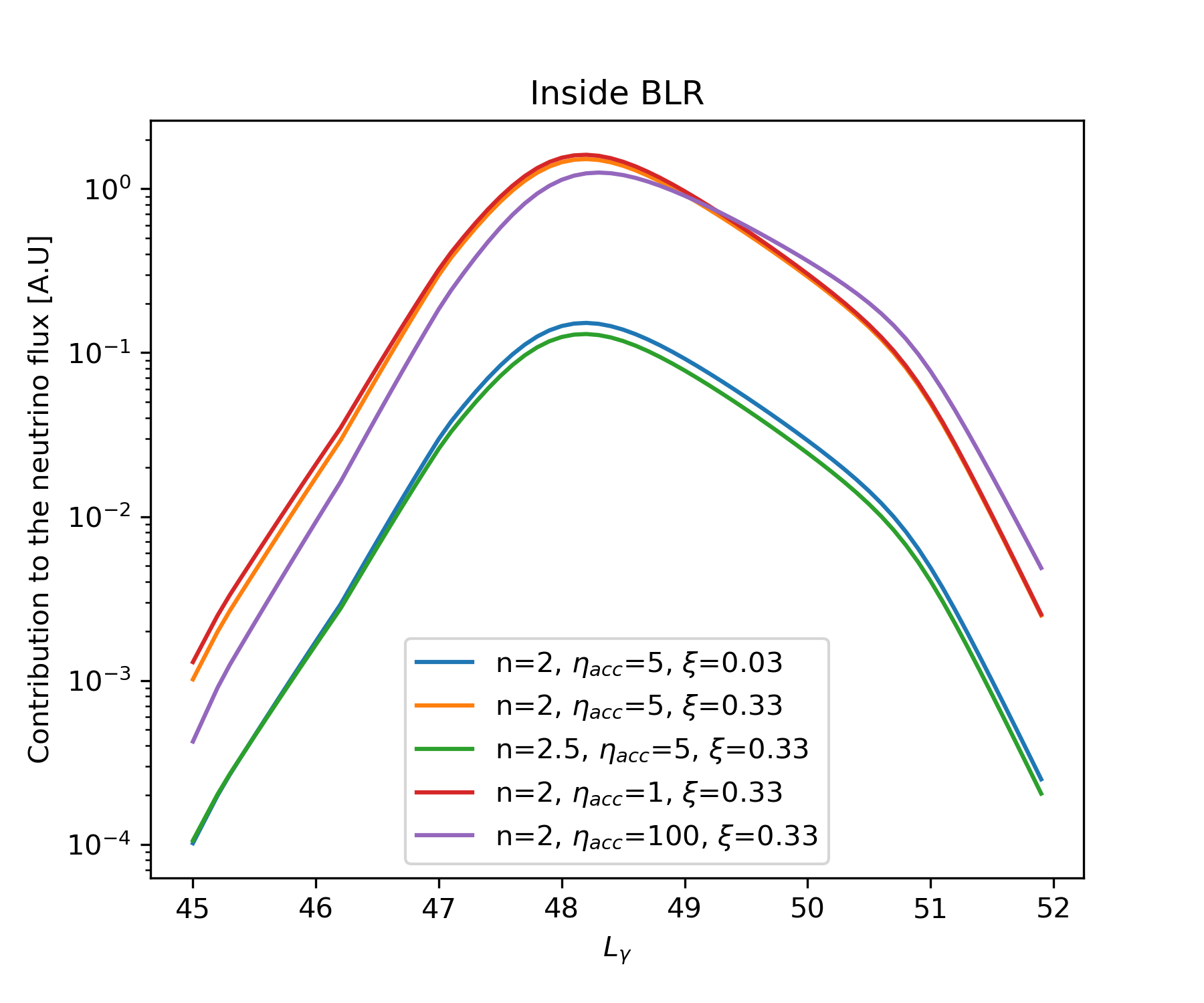}
\includegraphics[width=0.33\textwidth]{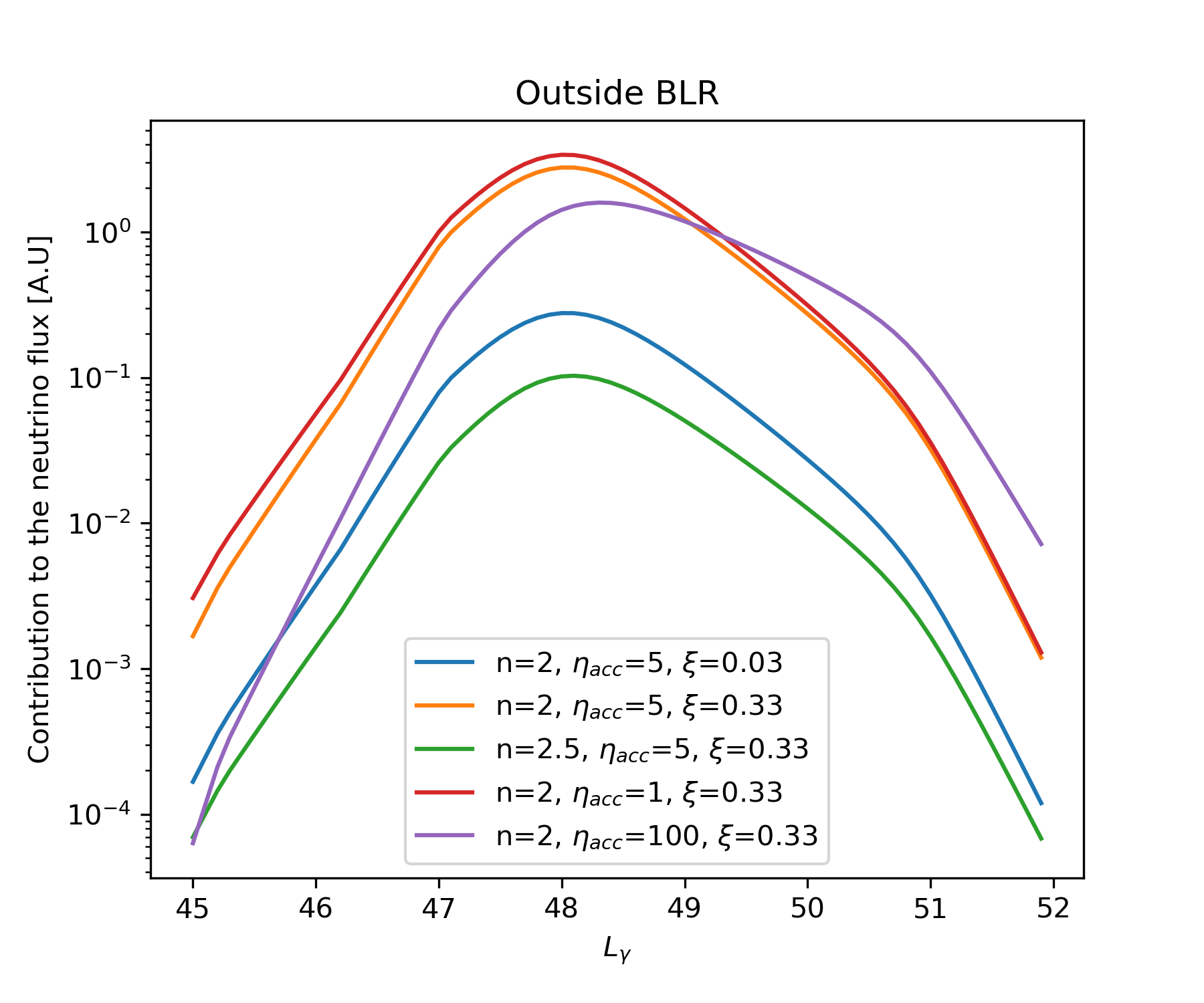}
\caption{{\it Left panel}: cosmological evolution of FSRQs, according to \protect\cite{Ajello:2011zi}.{\it Center and right panels}: Contribution to the neutrino flux for different parameters we considered in this work. Central panel refers to the ``inside'' BLR scenario, while right one, the ``outside'' scenario.}
\label{fig:ajello}
\end{figure*}

\subsection{Population}

To characterize the  entire population of FSRQs, we consider the results of \cite{Ajello:2011zi}. They offer a parametrization of the cosmological evolution of FSRQs, in terms of their gamma-ray luminosity, redshift and spectral index of the spectrum between 0.1-100 GeV. In our model we are only interested to the distribution in redshift and luminosity, therefore the spectral index dependence is integrated out. According to \cite{Ajello:2011zi} these high luminosity objects are characterized by a positive evolution and it becomes very strong for the most luminosity FSRQs, that are distributed as $\sim (1+z)^5$ for $z<1$. These objects are quite rare in the Universe, being characterized by a local density lower than 1 Gpc$^{-3}$. Moreover they are very brilliant varying in the range $L_\gamma=10^{44}-10^{50}$ erg/s.

In this work we divide the FSRQ population in 5 different luminosity bins: from $10^{44}$erg s$^{-1}<L_\gamma<10^{45}$erg s$^{-1}$ to $L_\gamma>10^{48}$erg s$^{-1}$. This classification is the same reported in G17 and it will be useful in the next section to derive the neutrino emission for each bin of FSRQs.
One possible realization of the distribution given in \cite{Ajello:2011zi} is shown in the first panel of Fig.\ref{fig:ajello}. In this plot we show the redshift and the gamma-ray luminosity of 1167 sources\footnote{The total number of sources comes directly from the theoretical distribution}. We also illustrate the separation between resolved and unresolved sources, using as a splitting point a luminosity at Earth of $4\times 10^{-12} \  \rm{erg \ cm^{-2} \ s^{-1} } $, similarly to \cite{Palladino:2018lov}.
In principle, one can expect that the bolometric (i.e. total) neutrino output is related to the power emitted in the electromagnetic channel. In \cite{MuraseWaxman16} or \cite{Righi:2016kio} for example authors assume a linear relation between the $\gamma$-ray luminosity and the neutrino one. Here we will find a non-linear relation between the two luminosities because of the models we considered (see section \ref{sec:models}).   

Center and right panels of fig.\ref{fig:ajello} show the expected contribution to the neutrino flux for different parameters studied in this work and for the two scenarios ``inside'' and ``outside'' BLR. In all the scenarios, high luminosity sources contribute most on the expected total neutrino flux from FSRQs.  

\section{Formalism for neutrino production}
\label{sec:neutrinos}
To derive the spectrum of the neutrino emission of FSRQ, we start from the (observed) total, or bolometric, electromagnetic non-thermal luminosity of the jet, $L_{\rm bol}$, and the (measured or inferred) disk luminosity, $L_{\rm d}$. The latter can be directly translated into the BLR and torus radii through Eq. \ref{eq:rblr} and \ref{eq:rtorus} above. This uniquely provide the external photon densities as a function of distance along the jet.

The other ingredient we need is the energy distribution of the relativistic cosmic rays (protons, for simplicity). We describe it as a power law with exponential cut-off (see below). We normalize the total proton luminosity (i.e. total energy injected into the proton population per unit time) as measured in the jet frame, $L^{\prime}_{\rm p}$ (primes denote quantities measured in the jet frame), to the electromagnetic luminosity, $L^{\prime}_{\rm rad}$, i.e. we impose $L^{\prime}_{\rm p}= \xi L^{\prime}_{\rm rad}$ (we assume this relation as a purely phenomenological description, although it can be justified by simple physical arguments, e.g. \citealt{Righi:2016kio}). The jet-frame electromagnetic luminosity can be derived from the observed luminosity with $L_{\rm bol}=L^{\prime}_{\rm rad} \delta^4$, where $\delta$ is the relativistic Doppler factor that depends on the the bulk Lorentz factor $\Gamma$ and the angle $\theta$ between the jet velocity and the line of sight as: $\delta=[\Gamma(1-\beta \cos \theta )]^{-1}$.
Using these relations we can finally write:
\begin{equation}
\label{eq:3}
L'_p=\frac{L_{\rm bol}\xi}{\delta^4}    
\end{equation}
For a given photomeson efficiency $f_{p\gamma}$, the luminosity channelled into neutrinos as measured in the jet frame is proportional to $f_{p\gamma}L'_p$. 

To derive the observed electromagnetic luminosity $L_{\rm bol}$ we exploit the averaged SED constructed for sources of different $\gamma$-ray luminosity by G17. Specifically, G17 consider a sample of 448 FSRQs with known redshift extracted from to the third catalog of AGN detected by \textit{Fermi}/LAT, 3LAC \citep{Ackermann15}, grouped them in 5 bins of gamma-ray luminosity and for each bin they derive an averaged SED using a suitable phenomenological function (see Eq. 11 of G17).

The SED models can be used to derive the average observed bolometric luminosity, $L_{\rm bol}$, of the five FSRQ categories. These luminosity bins are representative of the entire population of FSRQs.

We then calculate the neutrino spectrum adopting the same calculations described in \cite{Tavecchio14}. Here we describe schematically the main assumptions:

\begin{itemize}

\item The proton population is characterized by a total luminosity in the jet frame, $L'_{p}$ and the energy distribution is parametrised by a cut-offed power-law in energy:
\begin{equation}
\label{eq:4}
L'_p(E'_p)=k_pE_p^{'-n}\text{exp}\left(-\frac{E'_p}{E^{\prime}_{\rm cut}}\right) 
\end{equation}
\item The photomeson production efficiency $f_{p\gamma}$ is calculated as the ratio between the time-scales of the adiabatic losses ($t'_{\rm dyn}\approx R/c$, with $R$ is the radius of the jet region) and the cooling rate $t_{p\gamma}(E^{\prime}_p)$ of protons which is given, in the jet frame, by:
\begin{equation}
t_{p\gamma}^{-1}(E'_p)=c\int^{\infty}_{\epsilon_{\text{th}}/2\gamma^{\prime}_{\rm p}} d\epsilon \frac{n'_t(\epsilon)}{2\gamma'_p \epsilon^2} \int^{2\epsilon\gamma'_p}_{\epsilon_{\text{th}}} d\bar{\epsilon} \sigma_{p\gamma}(\bar{\epsilon})K_{p\gamma}(\bar{\epsilon})\bar\epsilon;
\label{eq:5}
\end{equation}
where $n_t$ is the numerical density of the target photons, $\gamma^{\prime}_p= E^{\prime}_p /m_pc^2$ and $\epsilon_{\text{th}}$ is the threshold energy of the process. We solve the integral of equation (\ref{eq:5}) using as cross section $\sigma_{p\gamma}$ and the inelasticity $K_{p\gamma}$ provided in \cite{AtoyanDermer}.

\item The neutrino luminosity $L'_\nu$ in the jet frame is given by (e.g., \citealt{Murase14}):
\begin{equation}
E'_\nu L'_\nu(E'_\nu)\simeq \frac{3}{8}f_{p\gamma}(E'_p)E'_p L'_p(E'_p); \qquad E'_\nu=0.05 E'_p
\end{equation}

\item From the decay of $\pi^0$ produced in the photomeson reaction, we expect a production of $\gamma$-rays with a luminosity $L'_\gamma$ given by:
\begin{equation}
\label{eq:gammafrompizero}
E'_\gamma L'_\gamma \simeq \frac{1}{2} f_{p\gamma}(E'_p)E'_p L'_p(E'_p); 
\end{equation}
\item Finally we obtain the neutrino luminosity $L_{\nu}$ in the observer frame using the Doppler factor $\delta$:
\begin{equation}
E_\nu L_{\nu}(E_\nu) = E'_\nu L'_\nu(E'_\nu )\delta^4; \qquad E_\nu = \delta E'_\nu.
\end{equation}
\end{itemize}

Summarizing, the free parameters entering in the calculation are: the index and the maximum energy of the proton distribution, $n$ and $E^{\prime}_{\rm cut}$, the distance $d$ of the region along the jet where the emission occurs, see Fig. \ref{fig:sketch_FSRQ}, the Lorentz factor $\Gamma$ and the Doppler factor $\delta$.
To reduce the number of free parameters, we fixed the Lorentz factor $\Gamma=13$ following  \cite{GGFT15} and we assume $\delta=\Gamma$ (i.e. we fix the observing angle to $1/\Gamma$).

\section{Models}
\label{sec:models}

In the following we consider two different regions inside the jet in which the photopion production occurs. In the first case (``inside BLR'') we assume that the region is at a distance $d<R_{\rm BLR}$ (where the BLR radius is given by eq. \ref{eq:rblr}) and, for definiteness, we will consider $d= R_{\rm BLR}/2$. In this case there are three different photon target populations involved in the $p\gamma$ reaction: the synchrotron radiation produced inside the jet (assumed to be cospatial with the protons), the radiation from the BLR clouds and the thermal torus radiation. In the second case (``outside BLR'') we assume the region to be beyond the BLR radius: in this case there are only the synchrotron radiation and the torus emission as photon targets for the $p\gamma$ reaction (the BLR radiation is strongly de-amplified in the jet frame, see e.g. \citealt{GGTF09}). In this case we fix $d=R_{\rm Torus}/2$ with $R_{\rm Torus}$ given by Eq. \ref{eq:rtorus}. 

$R_{\rm BLR}$ and $R_{\rm Torus}$ (and, therefore, the distance of the emission region in the two scenarios) can be derived from the disk luminosity. We link the disk luminosity to the bolometric non-thermal luminosity of the jet following \cite{GGTF09} (see also \citealt{GGNature}), i.e. we assume $L_{\rm d} \approx L_{\rm bol}/\Gamma^2$.

Assuming a jet semi-aperture angle $\theta_{\rm j}$, the radius of the emission region in the ``inside BLR'' case is:
\begin{equation}
R=\theta_{\rm j} d = \frac{R_{\rm BLR}}{2\Gamma}
\end{equation}
where we set the aperture angle to the standard value $\theta_{\rm j}=1/\Gamma$.
For the case outside the BLR we use the same equation with $R_{\rm Torus}$ instead of $R_{\rm BLR} $. 
In this way, we have fixed the distance and the size of the emission region, $d$ and $R$, to $L_{\rm bol}$ for each bin. 

Concerning the maximum energy of the proton distribution, $E^{\prime}_{\rm cut}$, we derive it using the condition of equality between cooling and acceleration time-scales $t'_{\rm cool} \approx t'_{\rm acc}$ in the jet frame. This condition corresponds to the equilibrium between losses and gains suffered by relativistic protons in the emission region. Losses include  photomeson losses with timescale $t'_{p\gamma}$, eq. \ref{eq:5} and adiabatic losses with characteristic timescale 
\begin{equation}
    t'_{\rm dyn} \approx \frac{R}{c}
\end{equation} 
with $R$ is the radius of the emission region (e.g. \citealt{Tavecchio15}).\\
The acceleration time can be expressed by:
\begin{equation}
\label{eq:10}
    t'_{\rm acc}(E'_p)=\eta_{\rm acc} \frac{r_L}{c} \simeq 1.36 \times 10^2 \eta_{\rm acc} \frac{E^{\prime}_{15}}{B} \;{\rm s}
\end{equation}
where $E^{\prime}_{15}$ is the proton energy normalized to $10^{15}$ eV, $B$ is the magnetic field (for definiteness we set $B=5$ G), and $\eta_{\rm acc}$ is a parameter depending on the details of the acceleration process and it is estimated to be in the range of 1-100 (e.g. \citealt{Rieger07}). For the acceleration time, the relevant parameter is $\eta_{\rm acc}/B$; then our scenario in which we fix the magnetic field but we keep $\eta_{\rm acc}$ a free parameter, is not affecting the calculations.

\begin{figure}
    \centering
    \includegraphics[scale=0.43]{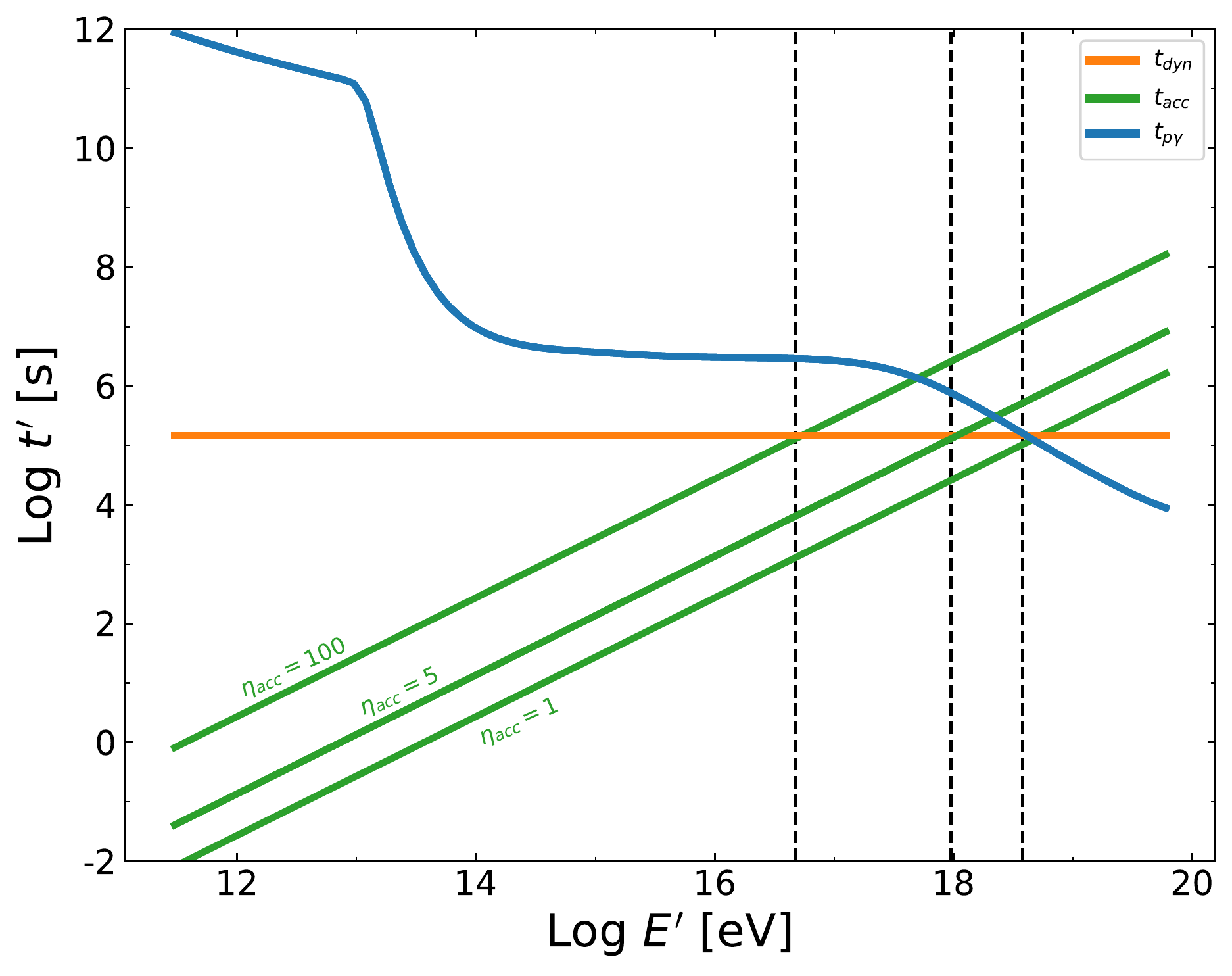}
    \caption{Acceleration and cooling time-scales (measured in the jet frame) expected for high-energy protons for an intermediate luminosity bin, $10^{46}$erg s$^{-1}<L_\gamma<10^{47}$erg s$^{-1}$. The blue line shows the photomeson time-scale $t_{p\gamma}$. Green lines show an estimate of the acceleration time-scale for three values of the acceleration efficiency $\eta_{\rm acc}$. The orange horizontal line is the adiabatic time-scale. All quantities are expressed in the emission region reference frame.}
    \label{fig:time}
\end{figure}

Figure \ref{fig:time} shows the relevant cooling (blue and orange lines) and acceleration time-scales (green lines) for different values of $\eta_{\rm acc}$ for an intermediate luminosity bin, $10^{46}$erg s$^{-1}<L_\gamma<10^{47}$erg s$^{-1}$. The intersection between the curve corresponding to the shorter cooling timescale and that for the acceleration timescale fixes the maximum energy attainable by accelerated protons.
The vertical black dashed lines indicates to the value of $E^{\prime}_{\rm cut}$ for different values of $\eta_{\rm acc}$. Except for the highest luminosity bins, the cooling is dominated by the adiabatic losses (horizontal orange line). 

\begin{figure*}
    \centering
    \includegraphics[scale=0.4]{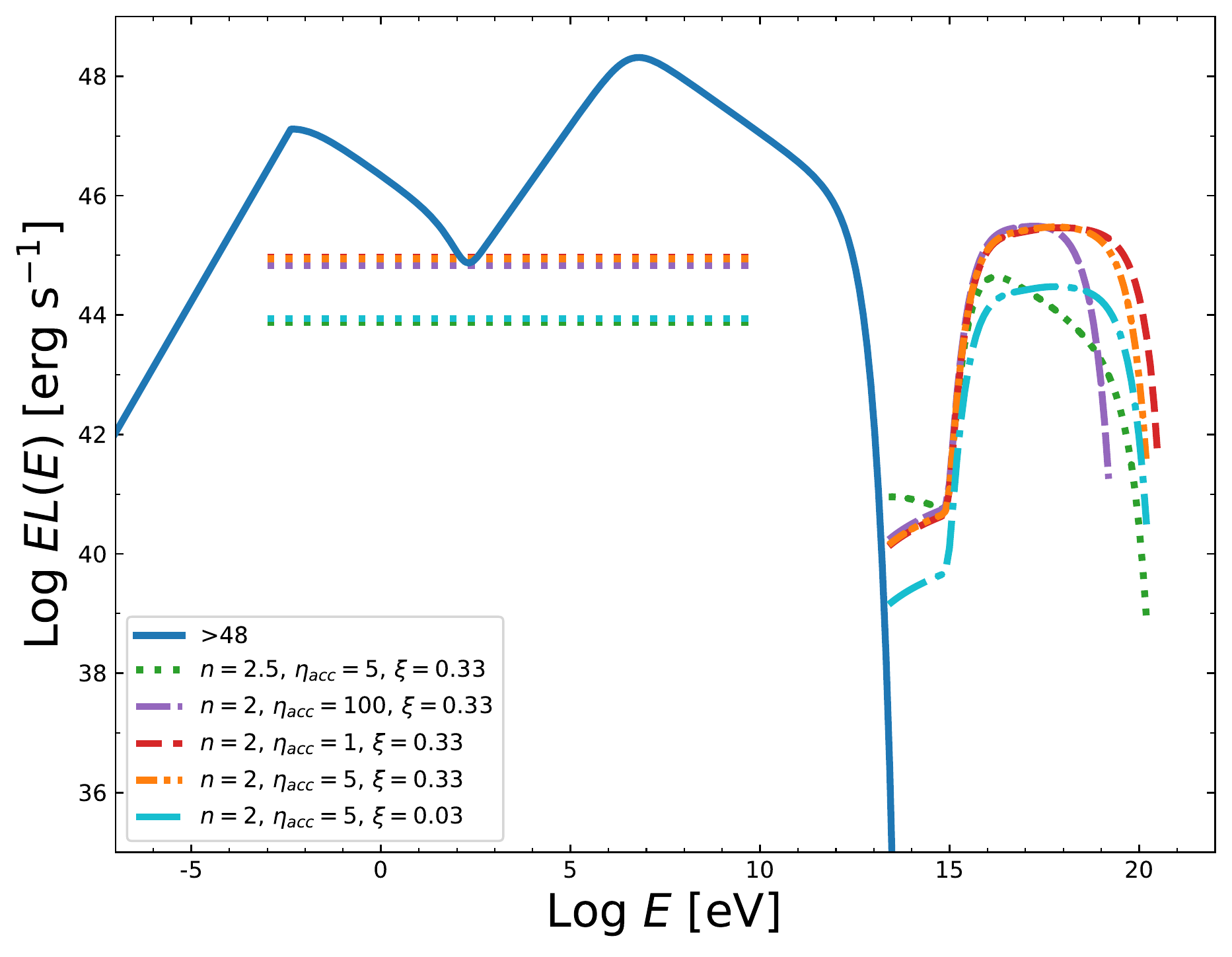}
    \includegraphics[scale=0.4]{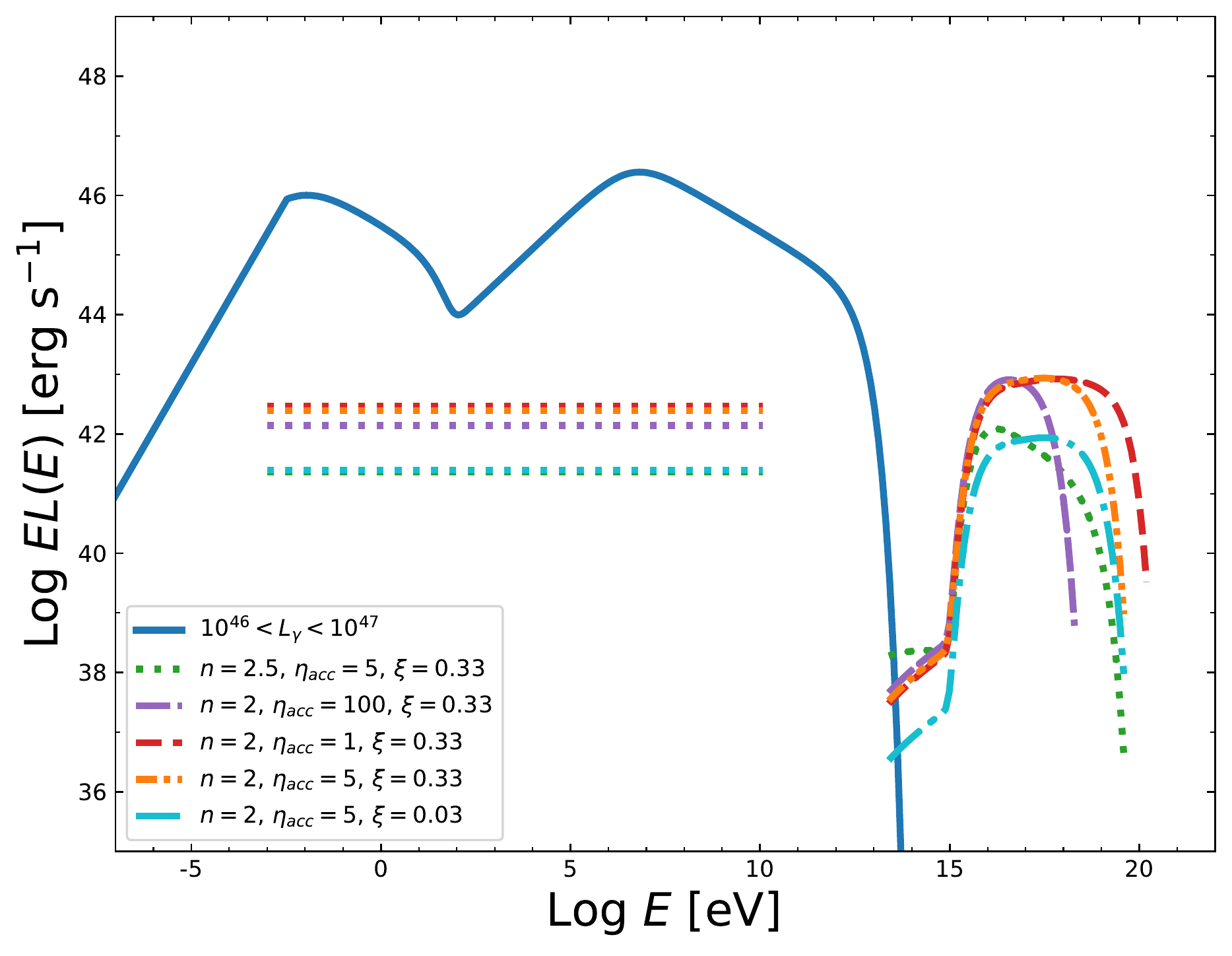}
    \caption{SED and neutrino component for an intermediate bin of FSRQ, $10^{46}$erg s$^{-1}<L_\gamma<10^{47}$erg s$^{-1}$, left panel, and the highest bin $L_\gamma>10^{48}$erg s$^{-1}$. Blue line shows the SED derived by G17. Dashed lines correspond to the expected cascade of $\gamma$-rays from the $\pi^0$ decay. In both cases we report the ``inside the BLR'' case.}
    \label{fig:SED46_pessimistic}
\end{figure*}

In this way the free parameters are now $\xi$ (the parameter linking $L_{\rm bol}$ with the cosmic ray luminosity, eq. \ref{eq:3}), $n$ the index of the proton energy distribution (eq. \ref{eq:4}) and $\eta_{\rm acc}$, the acceleration efficiency (eq. \ref{eq:10}). Fig. \ref{fig:SED46_pessimistic} shows how the neutrino spectrum changes for different values of $n$ and $\eta_{\rm acc}$ and two luminosity bins of FSRQs (the highest one $L_\gamma>10^{48}$erg s$^{-1}$, left panel, and the intermediate one, $10^{46}$erg s$^{-1}<L_\gamma<10^{47}$erg s$^{-1}$, right panel) considering the scenario inside the BLR. 

In Fig. \ref{fig:SED46_pessimistic} is clearly visible that the bulk of the neutrino output lies at energies above 1 PeV for the ``inside the BLR case'' and at even larger energies in the other case. This is a combined effect of the threshold and of the spectrum of the target photons. For instance, in the case in which the BLR photon field is dominant, as already remarked, the spectrum in the jet frame can be well approximated with a black body (BB) shape. Since protons interact mainly with photons with the energy peak of the BB spectrum, that means photons with an energy of
$\epsilon^{\prime}_{\rm BLR}\simeq 10 \, \Gamma \approx 130$ eV, by using the rule-of-thumb, only protons with $E^{\prime}_{\rm p}>10^{15}$ eV can interact. The neutrino produced through photomeson reactions will have an energy $E^{\prime}_{\nu}> 5\times 10^{13}$ eV and, in the observer frame,  $E_{\nu}=E^{\prime}_{\nu}\delta > 5\times 10^{13}\delta \approx 10^{15}$ eV. When IR photons from the torus are dominant, the threshold imposes an even larger energy for protons and, in turn, for the produced neutrinos. We remark that this result is completely independent on the details of the proton energy distribution (slope, maximum energy) but it is solely the result of the spectral properties of the photons fields.

\subsection{Limits from electromagnetic cascades}

As discussed in the previous paragraph, photomeson reactions produce neutrinos from the decay of charged pions and $\gamma$-rays from the $\pi^0$ decay. We can derive the spectrum of these photons using eq. \ref{eq:gammafrompizero}. Photons with these energies interacting with low-energy targets are promptly absorbed through the pair production reaction $\gamma\gamma\to e^+ + e^-$. The resulting pairs produce high-energy gamma-rays through inverse Compton scattering, triggering the development of an electromagnetic cascade inside the emission region. When the energy is degraded below the energy $E_{\tau=1}$ for which the optical depth is unity, photons can leave the source. The observed spectrum for saturated cascades is characterized by a universal spectrum with slope $\approx 1$ (i.e. ``flat'' in the SED representation) extending from a low energy end $E_{\rm min}$ to $E_{\tau=1}$.

The flat cascade component can provide a substantial contribution in the ``valley'' between the two SED bumps (as  extensively discussed for the case of TXS 0506+056, e.g. \citealt{MAGIC18}, \citealt{Cerruti18}). The very well know observational fact of the absence of features associated to cascades in the observed SED of FSRQ (e.g. G17) therefore implies a robust upper limit to the luminosity of the cascade component and, in turn, to the neutrino emission. 

A detailed calculation of the expected cascade spectrum is beyond our aims. For our purposes a good approximation of the cascade luminosity can be obtained assuming that the total UHE photon luminosity $L^{\prime}_{\gamma}$ injected by the $\pi^0$ decay (which in the observer frame is $L^{\prime}_{\gamma}\delta^4$) is re-emitted as a flat $\propto E^{-1}$ component extending from $E_{\rm min}$ to $E_{\tau=1}$. The latter energy can be derived from the condition that the optical depth of the emission region is $\tau(E)=1$.
$E_{\rm min}$ is more difficult to assess, but the normalization of the cascade spectrum has a very weak dependence on it. For definiteness we assume $E_{\rm min}=10^{-3}$ eV.

As an example, in fig. \ref{fig:SED46_pessimistic} we report (horizontal dotted lines) the cascade spectra derived for two luminosity bins and a set of parameters, together with the corresponding SED. For the high-luminosity bin ($L_\gamma>10^{48}$ erg s$^{-1}$, left panel) the cascade components are relatively more luminous than those corresponding to the lower luminosity bin ($10^{46}$erg s$^{-1}<L_\gamma<10^{47}$erg s$^{-1}$, right panel). Therefore, each set of parameters ($\eta_{\rm acc}$, $n$, $\xi$) fulfilling the condition that the cascades lie below the ``valley'' for the most powerful sources, automatically satisfies the constraint for the entire population. We define this the {\it low} model. 

Another possibility would be instead to adopt different sets of parameters for the different luminosity bins, tuning them so that for each bin the cascade provide the maximum contribution to neutrinos allowed by the SED. Clearly, this case describes the highest total neutrino flux allowed for the entire FSRQ population ({\it high} case). 
For simplicity, in this scenario we fixed the power index $n=2$ and the acceleration efficiency $\eta_{\rm acc}=5$. What we obtain is therefore a different value of $\xi$ for each luminosity bin of FSRQs in a range between $9 \times 10^{-1}<\xi<2\times 10^3$. This implies that low power FSRQs should be relatively more efficient than the high power ones in injecting power on relativistic protons.

\begin{figure*}
    \centering
    \includegraphics[width=0.45\textwidth]{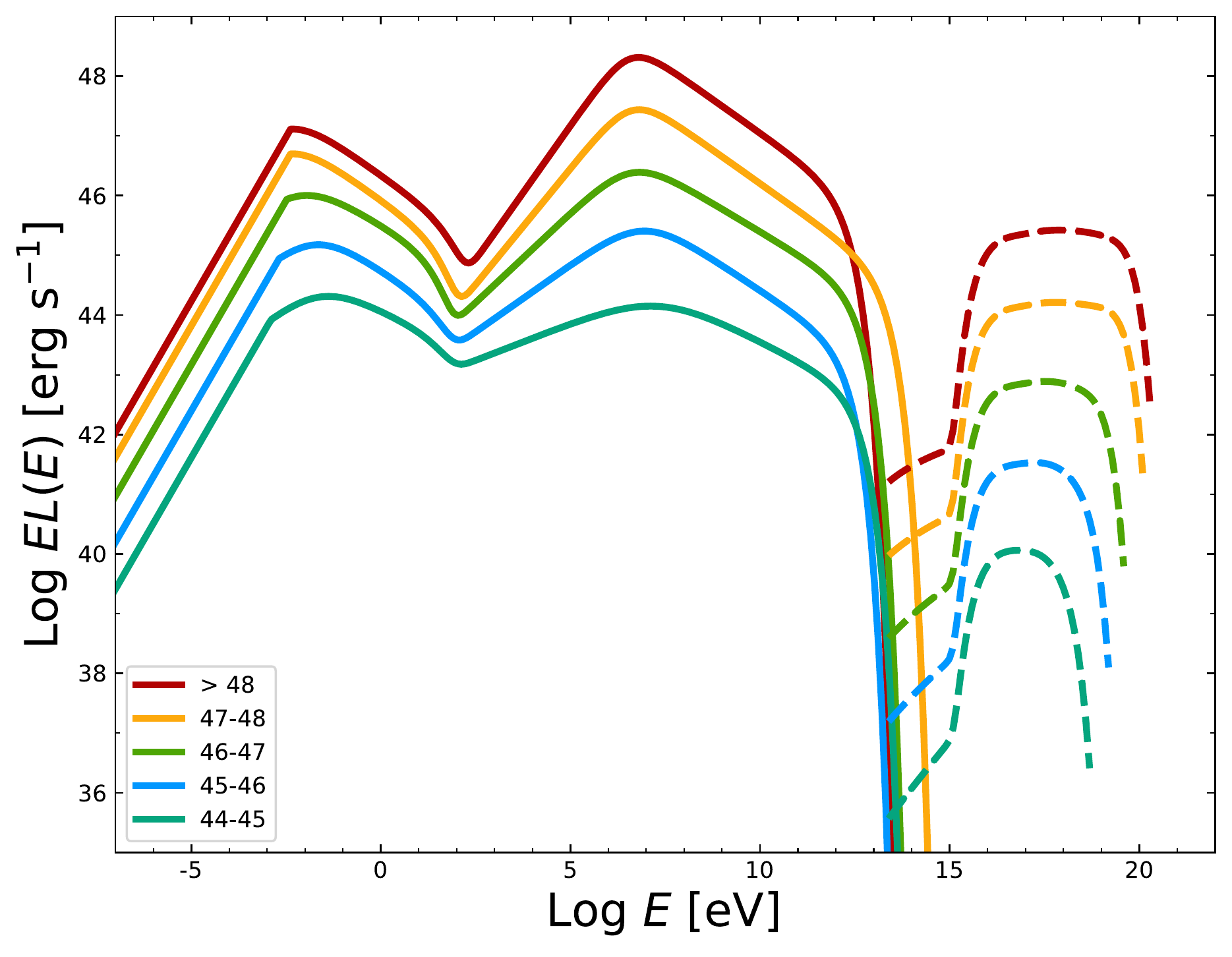}
    \includegraphics[width=0.45\textwidth]{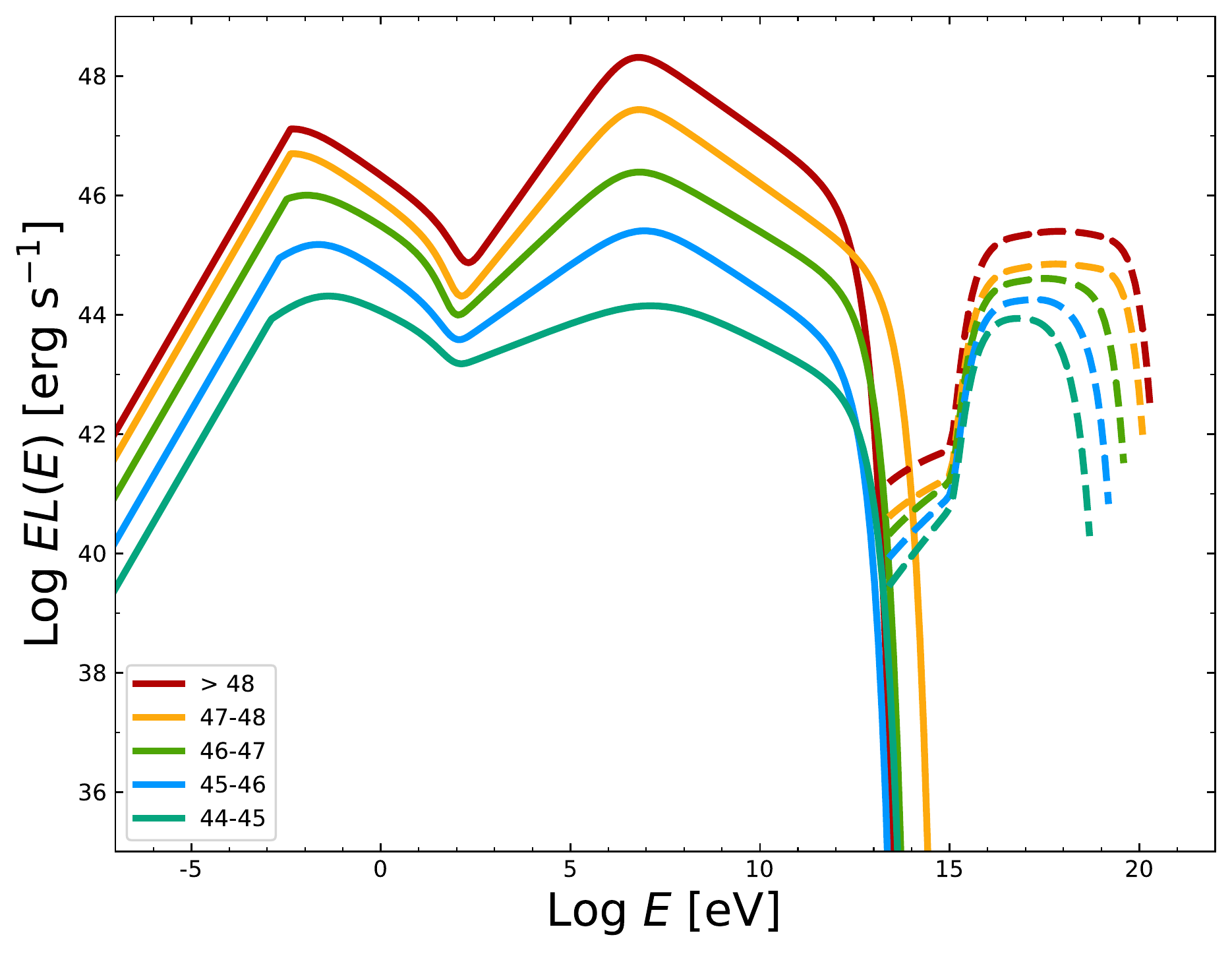}
    \caption{Electromagnetic SED of all categories of FSRQs, solid lines, with the expected neutrino spectra, dashed lines. \textit{Left panel:} Scenario with $n=2$, $\eta_{\rm acc}=5$ and $\xi=0.3$. \textit{Right panel:} Maximum scenario with $n=2$, $\eta_{\rm acc}=5$ and $\xi$ variable for each luminosity bin. }
    \label{fig:ALL_SED}
\end{figure*}

To give a comparison between the two cases, ``high'' and ``low'', we plot the neutrino spectra for all luminosity bins (inside the BLR case) in figure \ref{fig:ALL_SED}. In both panels solid lines are the electromagnetic SED from G17, while dashed lines are the neutrino spectra. In the left panel, the neutrino spectra are obtained using the same values $n=2$, $\eta_{\rm acc}=5$ and $\xi=0.3$ far all bins (low case). The right panel instead shows the high scenario with the same values of $n$ and $\eta_{\rm acc}$, but with $\xi$ variable for each bin (as above). It is clear that in this high scenario, the low luminosity FSRQs are more efficient in producing neutrinos than for the low case. 

A remark is in order. Even if the low case provides a neutrino flux lower than the maximum potentially possible (constrained by the cascade component), it should be considered a relatively optimistic scenario, since we are assuming that protons are accelerated and injected with a relatively large luminosity. Of course, an entire class of models with $\xi$ lower than the minimum assumed are possible, for which the predicted neutrino luminosity would therefore be lower than that derived here.

\section{Cumulative neutrino emission}
\label{sec:results}
\begin{figure*}
    \centering
    \includegraphics[width=0.45\textwidth]{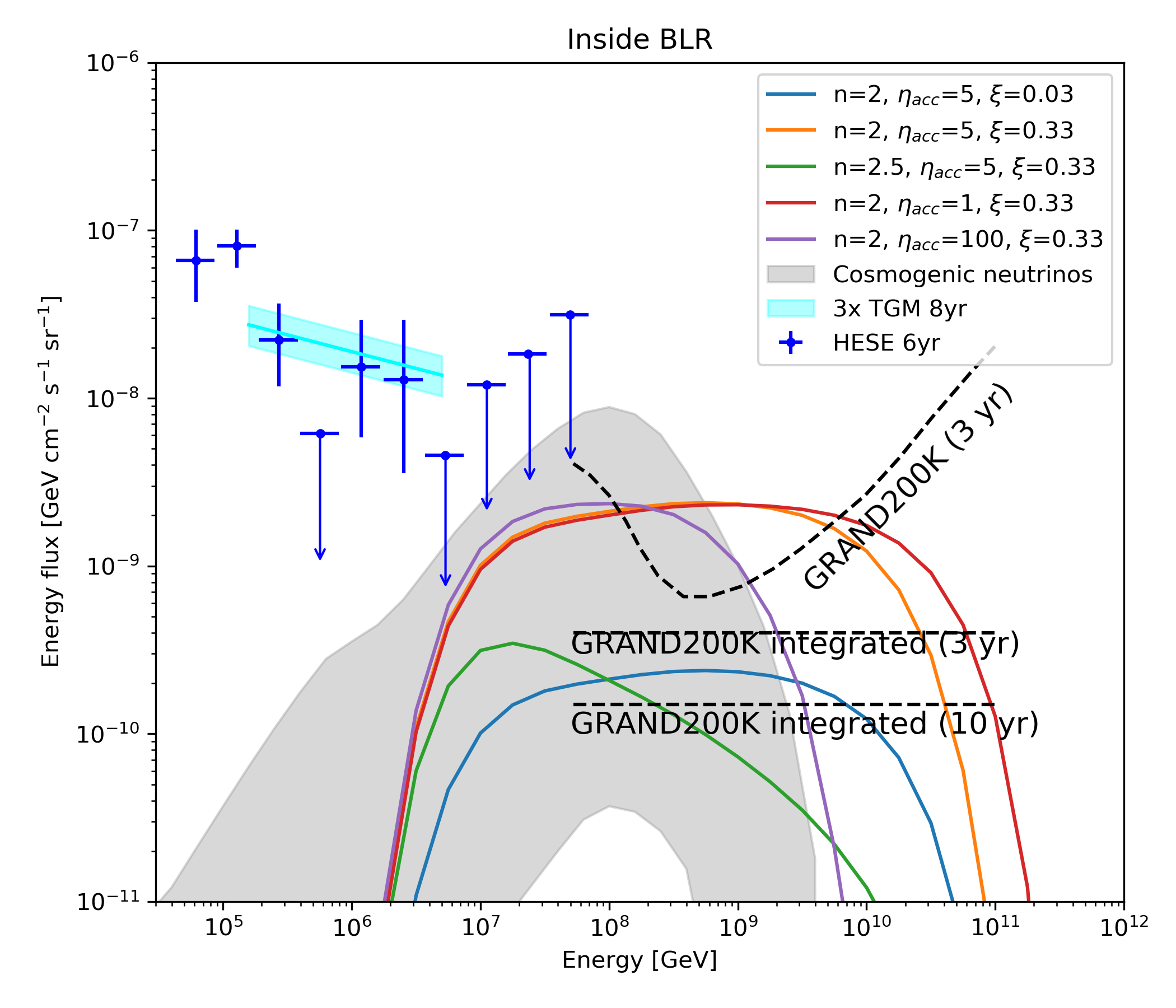}
    \includegraphics[width=0.45\textwidth]{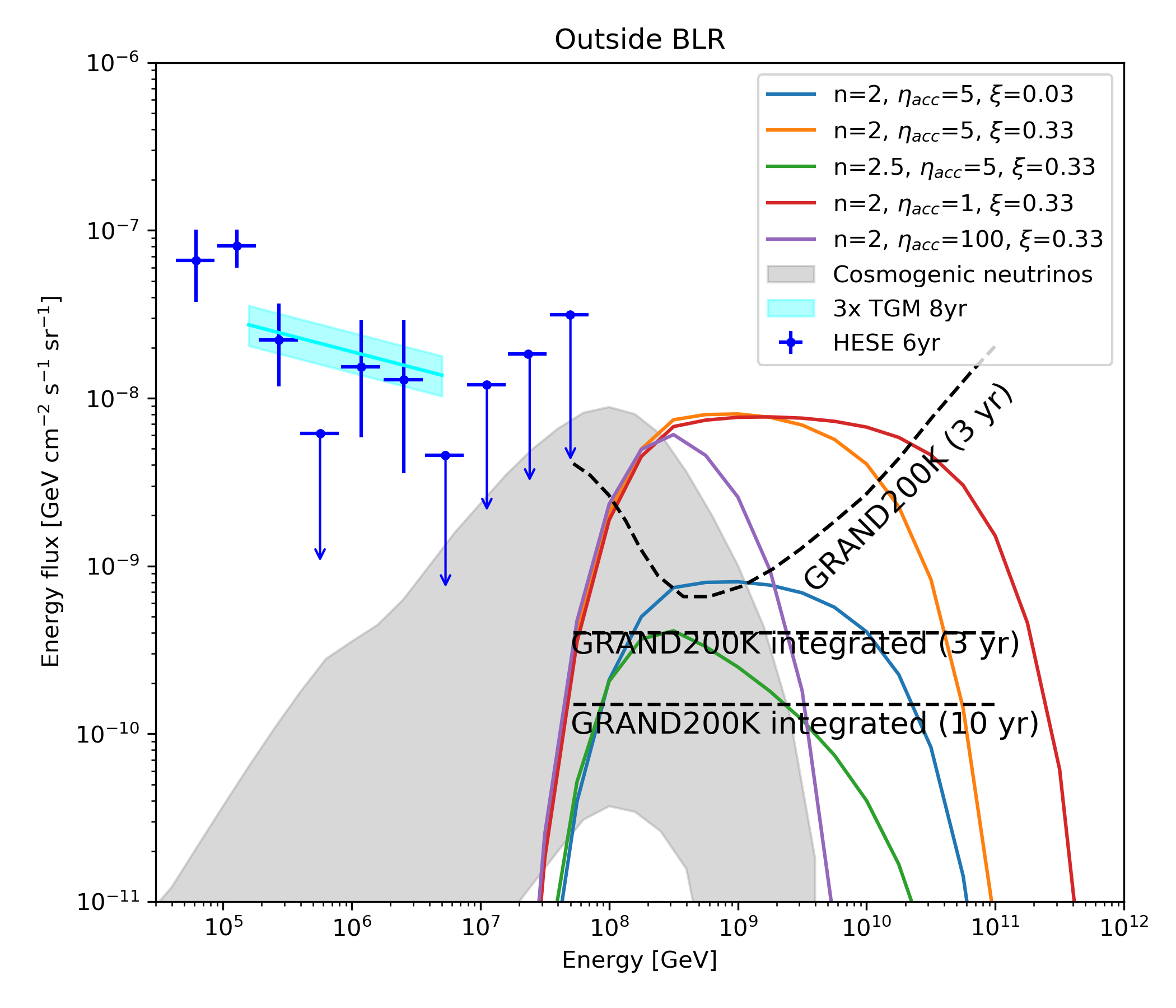}
    \caption{Diffuse neutrino case in the luminosity dependent case, assuming $\eta_{acc}=5$ in the left panel and $\eta_{acc}=50$ in the right panel. The results are compared with the IceCube data (HESE as blue points, throughgoing muons as light blue band) plus the sensitivity of the future experiment named GRAND.}
    \label{fig:diffusepessimistic}
\end{figure*}
\begin{figure}
\includegraphics[width=0.45\textwidth]{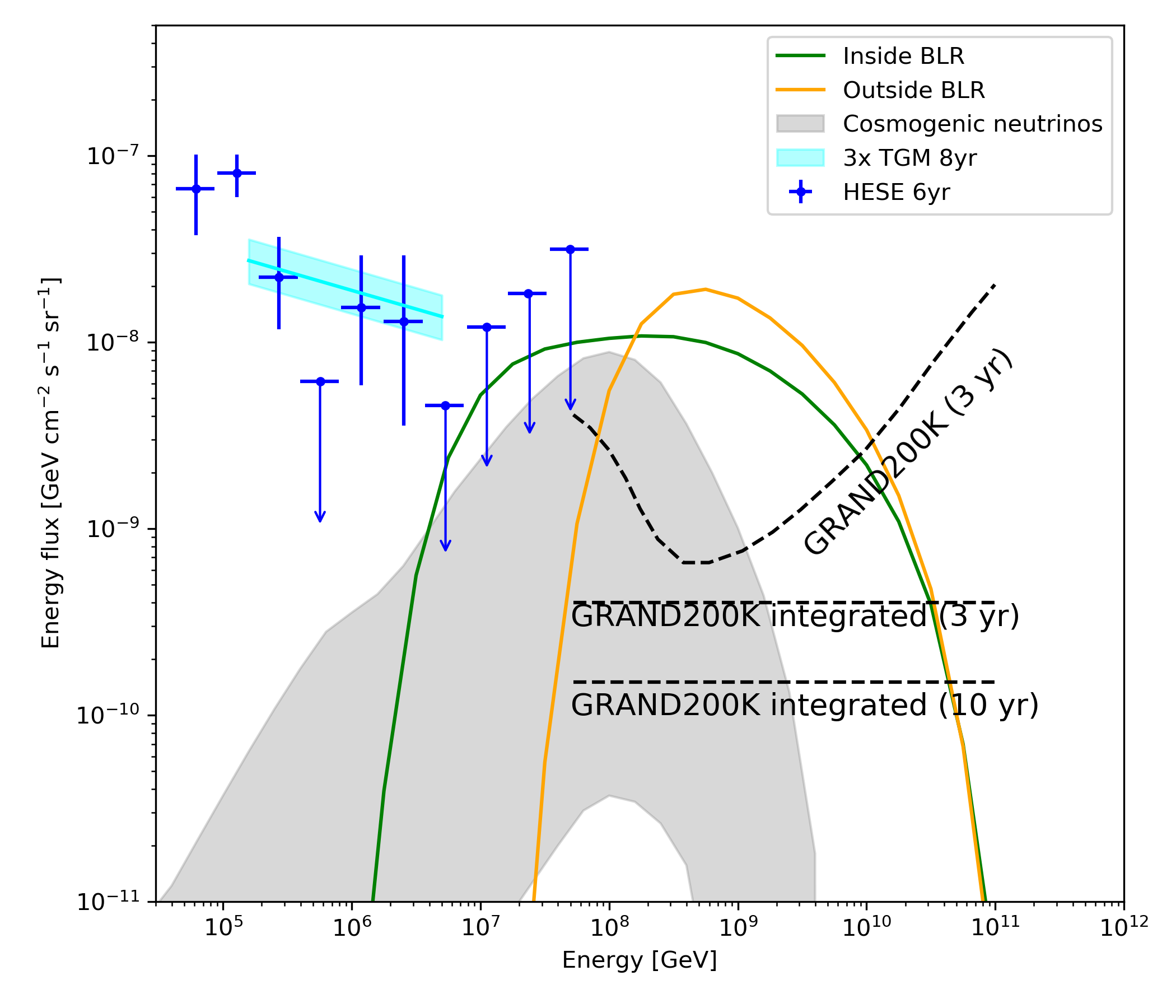}
\caption{Expected diffuse flux, assuming the maximum $\xi$ for each luminosity bin.}
\label{fig:optcase}
\end{figure}

Figures \ref{fig:diffusepessimistic} and \ref{fig:optcase} show the cumulative diffuse neutrino emission from FSRQs for the different scenarios that we consider in this work obtained by convolving the neutrinos spectra with the luminosity functions. In all cases we compare our results with the IceCube data (blue points for HESE events and light blue band for throughgoing muons, \citealt{Aartsen:2017mau}), the sensitivity of the future experiment GRAND, black dashed lines \citep{GRAND}, and the expected flux from cosmogenic neutrinos, gray region \citep{cosmogenicspectra}. Those neutrinos are produced by the interaction of ultra-high-energy cosmic rays, UHECR, interacting with the intergalactic backgrounds including the cosmic microwave background (CMB) and the extragalactic background light (EBL) \citep{Berez}.
The expected spectrum of the cosmogenic neutrinos displays a tail at low energies due to the interaction of cosmic rays with the UV component of EBL. In fact, the contribution of the interaction of UHECR with EBL is below $10^{18}$ eV, while at higher energies, the contribute of UHECR interacting with CMB dominates. 
In fig.\ref{fig:diffusepessimistic} we report our results considering the ``inside'' (left panel) and ``outside'' (right panel) scenarios in the ``low'' case.  We predicted neutrino emission from 10 different models (5 ``inside'' BLR, 5 ``outside'' BLR).
All scenarios are compatible with the current limits given by IceCube. 

In the left panel of \ref{fig:diffusepessimistic}, in which we are considering the emission region inside the BLR, we have two cases in which the maximum plateau of the emission by FSRQs is observable with 3 years of observation by GRAND and it is not interfering with the peak of the cosmogenic neutrinos (best expectation). Moreover, these two scenarios, in which we have respectively $n=2$, $\xi=0.3$ and $\eta_{\rm acc}=1$ (red solid line) or $\eta_{\rm acc}=5$ (yellow solid line), are the only two scenarios, among those we considered, providing fluxes clearly detectable with only 3 years with GRAND. For other scenarios (considering softer proton spectra, smaller $\xi$ or low acceleration efficiency) we predict spectra which are amidst in the cosmogenic neutrino region and more years are needed to clearly identified them.

In the right panel of fig. \ref{fig:diffusepessimistic} we show the ``outside'' BLR scenario. In this case the predicted spectra  peak at higher energies than the cosmogenic neutrino emission, because of the low-energy target photon energy from the torus. For this reason, we expect an easily distinction between the cosmogenic component and FSRQ neutrino emission. In addition to, we expect an isotropic flux of cosmogenic neutrinos, while we expect that neutrinos from FSRQs point back to the sources.

In figure \ref{fig:optcase}, the neutrino emission in case of the high scenario for both ``inside'' (green solid line) and ``outside'' (yellow line) are shown. In the first case the cosmogenic neutrino peak is below our prediction, while in the ``outside'' case, our spectrum is peaking at higher energy, allowing a possible detection of the cosmogenic neutrinos in case of the most optimistic scenario.  

In view of the physics programs of existing and future neutrino telescopes, this expectation deserves to be discussed carefully and quantitatively.



\section{Conclusion}
\label{sec:discussion}
Flat Spectrum Radio Quasars (FSRQs) are the most luminous blazars in the $\gamma$-ray band. Their nuclear region, that is naturally rich of photons, is the ideal environment to produce high energy neutrinos via photohadronic interactions. 
In this work we compute the neutrino flux expected from FSRQs, using standard assumptions on their structure and trying to use a limited set of the parameters.

In order to estimate the diffuse flux, we use the FSRQs population (i.e. luminosity function and evolution) described by \cite{Ajello:2011zi}. We considered two different scenarios, in which neutrinos are produced inside the broad line region (BLR), with three different photon target populations for the photomeson reaction, or outside the BLR, where only synchrotron and torus radiation contribute to the neutrino production. In both cases we derive the neutrino emission investigating 5 different scenarios in which the three main free parameters, the spectral index of relativistic protons $n$, the acceleration efficiency $\eta_{\rm acc}$ and the parameters $\xi$ that links the bolometric luminosity to the proton luminosity, are varying. 

The neutrino luminosity is constrained by the photons produced through the decay of $\pi^0$ that trigger an electromagnetic cascade. The resulting reprocessed photon emission is characterized by a flat spectrum extending over several decades in frequency, contributing especially in the X-ray band, i.e. in the region of the ``valley'' between the two SED peaks. The condition that the reprocessed flux lies below the observed (hard) X-ray continuum provides a stringent upper limit to the photomeson reactions. In particular, this provides a constraint on the parameters that we used in our luminosity independent model. In particular, the highest luminosity bin provides the most stringent constraint.

Then we considered a more complex model for both ``inside'' and ``outside'' scenario, in which all the luminosity bins of FSRQs are producing the maximum neutrino emission allowed without exceeding the limit given by the electromagnetic emission. In this model each luminosity bin had a different value of $\xi$ that implies that low-power FSRQs are relatively more efficient to inject power on relativistic protons than the high-power FSRQs. 

We remark that a possible dominant component of neutrinos well above 1 PeV is naturally related to the nature of the radiation fields filling the central regions of FSRQ. In fact, due to the threshold which characterizes the photopion reaction, only protons with energies above the PeV range (as measured in the jet frame) can interact and produce pions. Lower neutrino energies would require the existence of an unobserved intense field at soft/medium X-ray energies. This prediction is in agreement with the non-detection of FSRQ by the current generation of high-energy neutrino detectors.

We showed the cumulative neutrino flux for the entire population of FSRQ, and, in all scenarios we considered the FSRQ could be detectable by future neutrino telescopes in the subEeV-EeV energy region. Moreover, in several cases the neutrino flux is higher than the cosmogenic neutrino flux estimates discussed in the literature. Our results are in agreement with the results obtained in \cite{Rodrigues2017} in which authors identified high-luminous FSRQs as best neutrino emitters among blazar objects.


We would like to note that our results do not completely exclude the possibility that FSRQ provide some contribution to the main signal observed by IceCube. In fact, our analysis is based on the average spectra obtained from the sequence of FSRQ and assumes the structural properties described by the simple relations of Eqs.~\ref{eq:rblr}-\ref{eq:rtorus}. 
Therefore, our calculations should be considered  as  representative of the average population of FSRQ and do not describe possible outlier behaviours of the sources. Indeed, in some situations FSRQs could have different values of the various parameter that determine the neutrino emission, such as the Doppler factor or the power law index. In principle, similar modifications allow some special FSRQ - or most FSRQ in some special situation or state - to be more efficient  in  producing  neutrinos in  the  range $\sim 100$ TeV$-1$ PeV.


Our results can be easily tested using future neutrino telescope, since cosmogenic neutrinos are expected to be isotropically distributed, while EeV neutrinos from FSRQs should point back to the source, particularly to FSRQs having a gamma-ray luminosity between $10^{47}$ and $10^{49}$ erg/s.

\section*{Acknowledgements}
We would like to thank F. Halzen for his helpful suggestions, which helped us to improve the discussion of the results. 
AP has received funding from the European Research Council (ERC) under the European Union's Horizon 2020 research and innovation programme (Grant No. 646623). CR and FT acknowledge contribution from the grant INAF CTA–SKA, ``Probing particle acceleration and gamma-ray propagation with CTA and its precursors” and the INAF Main Stream project “High-energy extragalactic astrophysics: toward the Cherenkov Telescope Array''.
This work was partially supported by the research grant number 2017W4HA7S ``NAT-NET: Neutrino and Astroparticle Theory Network'' under the program PRIN 2017 funded by the Italian Ministero dell'Universit\`a e della Ricerca (MUR).

\bibliography{bibliography}

\begin{thebibliography}{51}
\providecommand{\natexlab}[1]{#1}
\providecommand{\url}[1]{\texttt{#1}}
\expandafter\ifx\csname urlstyle\endcsname\relax
  \providecommand{\doi}[1]{doi: #1}\else
  \providecommand{\doi}{doi: \begingroup \urlstyle{rm}\Url}\fi

\bibitem[Aartsen et~al.(2017)]{Aartsen:2017mau}
M.~G. Aartsen et~al.
\newblock {The IceCube Neutrino Observatory - Contributions to ICRC 2017 Part
  II: Properties of the Atmospheric and Astrophysical Neutrino Flux}.
\newblock 2017.

\bibitem[Aartsen et~al.(2018)]{IceCube:2018dnn}
M.~G. Aartsen et~al.
\newblock {Multimessenger observations of a flaring blazar coincident with
  high-energy neutrino IceCube-170922A}.
\newblock \emph{Science}, 361\penalty0 (6398):\penalty0 eaat1378, 2018.
\newblock \doi{10.1126/science.aat1378}.

\bibitem[{Ackermann} et~al.(2015){Ackermann}, {Ajello}, {Atwood}, {Baldini},
  {Ballet}, {Barbiellini}, {Bastieri}, {Becerra Gonzalez}, {Bellazzini},
  {Bissaldi}, {Bland ford}, {Bloom}, {Bonino}, {Bottacini}, {Brandt},
  {Bregeon}, {Britto}, {Bruel}, {Buehler}, {Buson}, {Caliandro}, {Cameron},
  {Caragiulo}, {Caraveo}, {Carpenter}, {Casandjian}, {Cavazzuti}, {Cecchi},
  {Charles}, {Chekhtman}, {Cheung}, {Chiang}, {Chiaro}, {Ciprini}, {Claus},
  {Cohen-Tanugi}, {Cominsky}, {Conrad}, {Cutini}, {D'Abrusco}, {D'Ammando}, {de
  Angelis}, {Desiante}, {Digel}, {Di Venere}, {Drell}, {Favuzzi}, {Fegan},
  {Ferrara}, {Finke}, {Focke}, {Franckowiak}, {Fuhrmann}, {Fukazawa},
  {Furniss}, {Fusco}, {Gargano}, {Gasparrini}, {Giglietto}, {Giommi},
  {Giordano}, {Giroletti}, {Glanzman}, {Godfrey}, {Grenier}, {Grove},
  {Guiriec}, {Hewitt}, {Hill}, {Horan}, {Itoh}, {J{\'o}hannesson}, {Johnson},
  {Johnson}, {Kataoka}, {Kawano}, {Krauss}, {Kuss}, {La Mura}, {Larsson},
  {Latronico}, {Leto}, {Li}, {Li}, {Longo}, {Loparco}, {Lott}, {Lovellette},
  {Lubrano}, {Madejski}, {Mayer}, {Mazziotta}, {McEnery}, {Michelson},
  {Mizuno}, {Moiseev}, {Monzani}, {Morselli}, {Moskalenko}, {Murgia}, {Nuss},
  {Ohno}, {Ohsugi}, {Ojha}, {Omodei}, {Orienti}, {Orland o}, {Paggi},
  {Paneque}, {Perkins}, {Pesce-Rollins}, {Piron}, {Pivato}, {Porter},
  {Rain{\`o}}, {Rando}, {Razzano}, {Razzaque}, {Reimer}, {Reimer}, {Romani},
  {Salvetti}, {Schaal}, {Schinzel}, {Schulz}, {Sgr{\`o}}, {Siskind},
  {Sokolovsky}, {Spada}, {Spandre}, {Spinelli}, {Stawarz}, {Suson},
  {Takahashi}, {Takahashi}, {Tanaka}, {Thayer}, {Thayer}, {Tibaldo}, {Torres},
  {Torresi}, {Tosti}, {Troja}, {Uchiyama}, {Vianello}, {Winer}, {Wood}, and
  {Zimmer}]{Ackermann15}
M.~{Ackermann}, M.~{Ajello}, W.~B. {Atwood}, L.~{Baldini}, J.~{Ballet},
  G.~{Barbiellini}, D.~{Bastieri}, J.~{Becerra Gonzalez}, R.~{Bellazzini},
  E.~{Bissaldi}, R.~D. {Bland ford}, E.~D. {Bloom}, R.~{Bonino},
  E.~{Bottacini}, T.~J. {Brandt}, J.~{Bregeon}, R.~J. {Britto}, P.~{Bruel},
  R.~{Buehler}, S.~{Buson}, G.~A. {Caliandro}, R.~A. {Cameron}, M.~{Caragiulo},
  P.~A. {Caraveo}, B.~{Carpenter}, J.~M. {Casandjian}, E.~{Cavazzuti},
  C.~{Cecchi}, E.~{Charles}, A.~{Chekhtman}, C.~C. {Cheung}, J.~{Chiang},
  G.~{Chiaro}, S.~{Ciprini}, R.~{Claus}, J.~{Cohen-Tanugi}, L.~R. {Cominsky},
  J.~{Conrad}, S.~{Cutini}, R.~{D'Abrusco}, F.~{D'Ammando}, A.~{de Angelis},
  R.~{Desiante}, S.~W. {Digel}, L.~{Di Venere}, P.~S. {Drell}, C.~{Favuzzi},
  S.~J. {Fegan}, E.~C. {Ferrara}, J.~{Finke}, W.~B. {Focke}, A.~{Franckowiak},
  L.~{Fuhrmann}, Y.~{Fukazawa}, A.~K. {Furniss}, P.~{Fusco}, F.~{Gargano},
  D.~{Gasparrini}, N.~{Giglietto}, P.~{Giommi}, F.~{Giordano}, M.~{Giroletti},
  T.~{Glanzman}, G.~{Godfrey}, I.~A. {Grenier}, J.~E. {Grove}, S.~{Guiriec},
  J.~W. {Hewitt}, A.~B. {Hill}, D.~{Horan}, R.~{Itoh}, G.~{J{\'o}hannesson},
  A.~S. {Johnson}, W.~N. {Johnson}, J.~{Kataoka}, T.~{Kawano}, F.~{Krauss},
  M.~{Kuss}, G.~{La Mura}, S.~{Larsson}, L.~{Latronico}, C.~{Leto}, J.~{Li},
  L.~{Li}, F.~{Longo}, F.~{Loparco}, B.~{Lott}, M.~N. {Lovellette},
  P.~{Lubrano}, G.~M. {Madejski}, M.~{Mayer}, M.~N. {Mazziotta}, J.~E.
  {McEnery}, P.~F. {Michelson}, T.~{Mizuno}, A.~A. {Moiseev}, M.~E. {Monzani},
  A.~{Morselli}, I.~V. {Moskalenko}, S.~{Murgia}, E.~{Nuss}, M.~{Ohno},
  T.~{Ohsugi}, R.~{Ojha}, N.~{Omodei}, M.~{Orienti}, E.~{Orland o}, A.~{Paggi},
  D.~{Paneque}, J.~S. {Perkins}, M.~{Pesce-Rollins}, F.~{Piron}, G.~{Pivato},
  T.~A. {Porter}, S.~{Rain{\`o}}, R.~{Rando}, M.~{Razzano}, S.~{Razzaque},
  A.~{Reimer}, O.~{Reimer}, R.~W. {Romani}, D.~{Salvetti}, M.~{Schaal}, F.~K.
  {Schinzel}, A.~{Schulz}, C.~{Sgr{\`o}}, E.~J. {Siskind}, K.~V. {Sokolovsky},
  F.~{Spada}, G.~{Spandre}, P.~{Spinelli}, L.~{Stawarz}, D.~J. {Suson},
  H.~{Takahashi}, T.~{Takahashi}, Y.~{Tanaka}, J.~G. {Thayer}, J.~B. {Thayer},
  L.~{Tibaldo}, D.~F. {Torres}, E.~{Torresi}, G.~{Tosti}, E.~{Troja},
  Y.~{Uchiyama}, G.~{Vianello}, B.~L. {Winer}, K.~S. {Wood}, and S.~{Zimmer}.
\newblock {The Third Catalog of Active Galactic Nuclei Detected by the Fermi
  Large Area Telescope}.
\newblock \emph{\apj}, 810\penalty0 (1):\penalty0 14, September 2015.
\newblock \doi{10.1088/0004-637X/810/1/14}.

\bibitem[{Aharonian}(2000)]{Aharonian00}
F.~A. {Aharonian}.
\newblock {TeV gamma rays from BL Lac objects due to synchrotron radiation of
  extremely high energy protons}.
\newblock \emph{New Astronomy}, 5:\penalty0 377--395, November 2000.
\newblock \doi{10.1016/S1384-1076(00)00039-7}.

\bibitem[{Ahlers} and {Halzen}(2014)]{Ahalers14}
M.~{Ahlers} and F.~{Halzen}.
\newblock {Pinpointing extragalactic neutrino sources in light of recent
  IceCube observations}.
\newblock \emph{Physical Review D}, 90\penalty0 (4):\penalty0 043005, August
  2014.
\newblock \doi{10.1103/PhysRevD.90.043005}.

\bibitem[Ajello et~al.(2012)]{Ajello:2011zi}
M.~Ajello et~al.
\newblock {The Luminosity Function of Fermi-detected Flat-Spectrum Radio
  Quasars}.
\newblock \emph{Astrophys. J.}, 751:\penalty0 108, 2012.
\newblock \doi{10.1088/0004-637X/751/2/108}.

\bibitem[{{\'A}lvarez-Mu{\~n}iz} et~al.(2020){{\'A}lvarez-Mu{\~n}iz}, {Alves
  Batista}, {Balagopal V.}, {Bolmont}, {Bustamante}, {Carvalho}, {Charrier},
  {Cognard}, {Decoene}, {Denton}, {De Jong}, {De Vries}, {Engel}, {Fang},
  {Finley}, {Gabici}, {Gou}, {Gu}, {Gu{\'e}pin}, {Hu}, {Huang}, {Kotera}, {Le
  Coz}, {Lenain}, {L{\"u}}, {Martineau-Huynh}, {Mostaf{\'a}}, {Mottez},
  {Murase}, {Niess}, {Oikonomou}, {Pierog}, {Qian}, {Qin}, {Ran},
  {Renault-Tinacci}, {Roth}, {Schr{\"o}der}, {Sch{\"u}ssler}, {Tasse},
  {Timmermans}, {Tueros}, {Wu}, {Zarka}, {Zech}, {Zhang}, {Zhang}, {Zhang},
  {Zheng}, and {Zilles}]{GRAND}
Jaime {{\'A}lvarez-Mu{\~n}iz}, Rafael {Alves Batista}, Aswathi {Balagopal V.},
  Julien {Bolmont}, Mauricio {Bustamante}, Washington {Carvalho}, Didier
  {Charrier}, Isma{\"e}l {Cognard}, Valentin {Decoene}, Peter~B. {Denton},
  Sijbrand {De Jong}, Krijn~D. {De Vries}, Ralph {Engel}, Ke~{Fang}, Chad
  {Finley}, Stefano {Gabici}, QuanBu {Gou}, JunHua {Gu}, Claire {Gu{\'e}pin},
  HongBo {Hu}, Yan {Huang}, Kumiko {Kotera}, Sand~ra {Le Coz}, Jean-Philippe
  {Lenain}, GuoLiang {L{\"u}}, Olivier {Martineau-Huynh}, Miguel {Mostaf{\'a}},
  Fabrice {Mottez}, Kohta {Murase}, Valentin {Niess}, Foteini {Oikonomou},
  Tanguy {Pierog}, XiangLi {Qian}, Bo~{Qin}, Duan {Ran}, Nicolas
  {Renault-Tinacci}, Markus {Roth}, Frank~G. {Schr{\"o}der}, Fabian
  {Sch{\"u}ssler}, Cyril {Tasse}, Charles {Timmermans}, Mat{\'\i}as {Tueros},
  XiangPing {Wu}, Philippe {Zarka}, Andreas {Zech}, B.~Theodore {Zhang}, JianLi
  {Zhang}, Yi~{Zhang}, Qian {Zheng}, and Anne {Zilles}.
\newblock {The Giant Radio Array for Neutrino Detection (GRAND): Science and
  design}.
\newblock \emph{Science China Physics, Mechanics, and Astronomy}, 63\penalty0
  (1):\penalty0 219501, January 2020.
\newblock \doi{10.1007/s11433-018-9385-7}.

\bibitem[{Alves Batista} et~al.(2019){Alves Batista}, {de Almeida}, {Lago}, and
  {Kotera}]{cosmogenicspectra}
Rafael {Alves Batista}, Rogerio~M. {de Almeida}, Bruno {Lago}, and Kumiko
  {Kotera}.
\newblock {Cosmogenic photon and neutrino fluxes in the Auger era}.
\newblock \emph{\jcap}, 2019\penalty0 (1):\penalty0 002, January 2019.
\newblock \doi{10.1088/1475-7516/2019/01/002}.

\bibitem[{Ansoldi} et~al.(2018){Ansoldi}, {Antonelli}, {Arcaro}, {Baack},
  {Babi{\'c}}, {Banerjee}, {Bangale}, {Barres de Almeida}, {Barrio}, {Becerra
  Gonz{\'a}lez}, {Bednarek}, {Bernardini}, {Berse}, {Berti}, {Besenrieder},
  {Bhattacharyya}, {Bigongiari}, {Biland}, {Blanch}, {Bonnoli}, {Carosi},
  {Ceribella}, {Chatterjee}, {Colak}, {Colin}, {Colombo}, {Contreras},
  {Cortina}, {Covino}, {Cumani}, {D'Elia}, {Da Vela}, {Dazzi}, {De Angelis},
  {De Lotto}, {Delfino}, {Delgado}, {Di Pierro}, {Dom{\'{\i}}nguez}, {Dominis
  Prester}, {Dorner}, {Doro}, {Einecke}, {Elsaesser}, {Fallah Ramazani},
  {Fattorini}, {Fern{\'a}ndez-Barral}, {Ferrara}, {Fidalgo}, {Foffano},
  {Fonseca}, {Font}, {Fruck}, {Gallozzi}, {Garc{\'{\i}}a L{\'o}pez},
  {Garczarczyk}, {Gaug}, {Giammaria}, {Godinovi{\'c}}, {Guberman}, {Hadasch},
  {Hahn}, {Hassan}, {Hayashida}, {Herrera}, {Hoang}, {Hrupec}, {Inoue},
  {Ishio}, {Iwamura}, {Konno}, {Kubo}, {Kushida}, {Lamastra}, {Lelas}, {Leone},
  {Lindfors}, {Lombardi}, {Longo}, {L{\'o}pez}, {Maggio}, {Majumdar},
  {Makariev}, {Maneva}, {Manganaro}, {Mannheim}, {Maraschi}, {Mariotti},
  {Mart{\'{\i}}nez}, {Masuda}, {Mazin}, {Mielke}, {Minev}, {Miranda},
  {Mirzoyan}, {Moralejo}, {Moreno}, {Moretti}, {Neustroev}, {Niedzwiecki},
  {Nievas Rosillo}, {Nigro}, {Nilsson}, {Ninci}, {Nishijima}, {Noda},
  {Nogu{\'e}s}, {Paiano}, {Palacio}, {Paneque}, {Paoletti}, {Paredes},
  {Pedaletti}, {Pe{\~n}il}, {Peresano}, {Persic}, {Pfrang}, {Prada Moroni},
  {Prandini}, {Puljak}, {Garcia}, {Rhode}, {Rib{\'o}}, {Rico}, {Righi},
  {Rugliancich}, {Saha}, {Saito}, {Satalecka}, {Schweizer}, {Sitarek}, {{\v
  S}nidari{\'c}}, {Sobczynska}, {Stamerra}, {Strzys}, {Suri{\'c}}, {Tavecchio},
  {Temnikov}, {Terzi{\'c}}, {Teshima}, {Torres-Alb{\'a}}, {Tsujimoto}, {Vanzo},
  {Vazquez Acosta}, {Vovk}, {Ward}, {Will}, {Zari{\'c}}, and
  {Cerruti}]{MAGIC18}
S.~{Ansoldi}, L.~A. {Antonelli}, C.~{Arcaro}, D.~{Baack}, A.~{Babi{\'c}},
  B.~{Banerjee}, P.~{Bangale}, U.~{Barres de Almeida}, J.~A. {Barrio},
  J.~{Becerra Gonz{\'a}lez}, W.~{Bednarek}, E.~{Bernardini}, R.~C. {Berse},
  A.~{Berti}, J.~{Besenrieder}, W.~{Bhattacharyya}, C.~{Bigongiari},
  A.~{Biland}, O.~{Blanch}, G.~{Bonnoli}, R.~{Carosi}, G.~{Ceribella},
  A.~{Chatterjee}, S.~M. {Colak}, P.~{Colin}, E.~{Colombo}, J.~L. {Contreras},
  J.~{Cortina}, S.~{Covino}, P.~{Cumani}, V.~{D'Elia}, P.~{Da Vela},
  F.~{Dazzi}, A.~{De Angelis}, B.~{De Lotto}, M.~{Delfino}, J.~{Delgado},
  F.~{Di Pierro}, A.~{Dom{\'{\i}}nguez}, D.~{Dominis Prester}, D.~{Dorner},
  M.~{Doro}, S.~{Einecke}, D.~{Elsaesser}, V.~{Fallah Ramazani},
  A.~{Fattorini}, A.~{Fern{\'a}ndez-Barral}, G.~{Ferrara}, D.~{Fidalgo},
  L.~{Foffano}, M.~V. {Fonseca}, L.~{Font}, C.~{Fruck}, S.~{Gallozzi}, R.~J.
  {Garc{\'{\i}}a L{\'o}pez}, M.~{Garczarczyk}, M.~{Gaug}, P.~{Giammaria},
  N.~{Godinovi{\'c}}, D.~{Guberman}, D.~{Hadasch}, A.~{Hahn}, T.~{Hassan},
  M.~{Hayashida}, J.~{Herrera}, J.~{Hoang}, D.~{Hrupec}, S.~{Inoue},
  K.~{Ishio}, Y.~{Iwamura}, Y.~{Konno}, H.~{Kubo}, J.~{Kushida}, A.~{Lamastra},
  D.~{Lelas}, F.~{Leone}, E.~{Lindfors}, S.~{Lombardi}, F.~{Longo},
  M.~{L{\'o}pez}, C.~{Maggio}, P.~{Majumdar}, M.~{Makariev}, G.~{Maneva},
  M.~{Manganaro}, K.~{Mannheim}, L.~{Maraschi}, M.~{Mariotti},
  M.~{Mart{\'{\i}}nez}, S.~{Masuda}, D.~{Mazin}, K.~{Mielke}, M.~{Minev}, J.~M.
  {Miranda}, R.~{Mirzoyan}, A.~{Moralejo}, V.~{Moreno}, E.~{Moretti},
  V.~{Neustroev}, A.~{Niedzwiecki}, M.~{Nievas Rosillo}, C.~{Nigro},
  K.~{Nilsson}, D.~{Ninci}, K.~{Nishijima}, K.~{Noda}, L.~{Nogu{\'e}s},
  S.~{Paiano}, J.~{Palacio}, D.~{Paneque}, R.~{Paoletti}, J.~M. {Paredes},
  G.~{Pedaletti}, P.~{Pe{\~n}il}, M.~{Peresano}, M.~{Persic}, K.~{Pfrang},
  P.~G. {Prada Moroni}, E.~{Prandini}, I.~{Puljak}, J.~R. {Garcia}, W.~{Rhode},
  M.~{Rib{\'o}}, J.~{Rico}, C.~{Righi}, A.~{Rugliancich}, L.~{Saha},
  T.~{Saito}, K.~{Satalecka}, T.~{Schweizer}, J.~{Sitarek}, I.~{{\v
  S}nidari{\'c}}, D.~{Sobczynska}, A.~{Stamerra}, M.~{Strzys}, T.~{Suri{\'c}},
  F.~{Tavecchio}, P.~{Temnikov}, T.~{Terzi{\'c}}, M.~{Teshima},
  N.~{Torres-Alb{\'a}}, S.~{Tsujimoto}, G.~{Vanzo}, M.~{Vazquez Acosta},
  I.~{Vovk}, J.~E. {Ward}, M.~{Will}, D.~{Zari{\'c}}, and M.~{Cerruti}.
\newblock {The Blazar TXS 0506+056 Associated with a High-energy Neutrino:
  Insights into Extragalactic Jets and Cosmic-Ray Acceleration}.
\newblock \emph{Astrophysical Journal, Letters}, 863:\penalty0 L10, August
  2018.
\newblock \doi{10.3847/2041-8213/aad083}.

\bibitem[{Atoyan} and {Dermer}(2003)]{AtoyanDermer}
Armen~M. {Atoyan} and Charles~D. {Dermer}.
\newblock {Neutral Beams from Blazar Jets}.
\newblock \emph{\apj}, 586\penalty0 (1):\penalty0 79--96, March 2003.
\newblock \doi{10.1086/346261}.

\bibitem[Berezinsky and Zatsepin(1969)]{Berez}
V.~S. Berezinsky and G.~T. Zatsepin.
\newblock {Cosmic rays at ultrahigh-energies (neutrino?)}.
\newblock \emph{Phys. Lett.}, 28B:\penalty0 423--424, 1969.
\newblock \doi{10.1016/0370-2693(69)90341-4}.

\bibitem[{Blandford} et~al.(2019){Blandford}, {Meier}, and
  {Readhead}]{BlandfordreviewAGN}
Roger {Blandford}, David {Meier}, and Anthony {Readhead}.
\newblock {Relativistic Jets from Active Galactic Nuclei}.
\newblock \emph{\araa}, 57:\penalty0 467--509, August 2019.
\newblock \doi{10.1146/annurev-astro-081817-051948}.

\bibitem[{Cerruti} et~al.(2018){Cerruti}, {Zech}, {Boisson}, {Emery}, {Inoue},
  and {Lenain}]{Cerruti18}
M.~{Cerruti}, A.~{Zech}, C.~{Boisson}, G.~{Emery}, S.~{Inoue}, and J.-P.
  {Lenain}.
\newblock {Lepto-hadronic single-zone models for the electromagnetic and
  neutrino emission of TXS 0506+056}.
\newblock \emph{ArXiv e-prints}, July 2018.

\bibitem[{Costamante} et~al.(2018){Costamante}, {Cutini}, {Tosti}, {Antolini},
  and {Tramacere}]{Costamante18}
L.~{Costamante}, S.~{Cutini}, G.~{Tosti}, E.~{Antolini}, and A.~{Tramacere}.
\newblock {On the origin of gamma-rays in Fermi blazars: beyondthe broad-line
  region}.
\newblock \emph{\mnras}, 477\penalty0 (4):\penalty0 4749--4767, July 2018.
\newblock \doi{10.1093/mnras/sty887}.

\bibitem[Dermer et~al.(2014)Dermer, Murase, and Inoue]{Dermer:2014vaa}
Charles~D. Dermer, Kohta Murase, and Yoshiyuki Inoue.
\newblock {Photopion Production in Black-Hole Jets and Flat-Spectrum Radio
  Quasars as PeV Neutrino Sources}.
\newblock \emph{JHEAp}, 3-4:\penalty0 29--40, 2014.
\newblock \doi{10.1016/j.jheap.2014.09.001}.

\bibitem[{Fossati} et~al.(1998){Fossati}, {Maraschi}, {Celotti}, {Comastri},
  and {Ghisellini}]{Fossati98}
G.~{Fossati}, L.~{Maraschi}, A.~{Celotti}, A.~{Comastri}, and G.~{Ghisellini}.
\newblock {A unifying view of the spectral energy distributions of blazars}.
\newblock \emph{Monthly Notices of the RAS}, 299:\penalty0 433--448, September
  1998.
\newblock \doi{10.1046/j.1365-8711.1998.01828.x}.

\bibitem[{Gaisser}(2018)]{Gaisser18}
T.~K. {Gaisser}.
\newblock {Neutrino Astronomy 2017}.
\newblock \emph{ArXiv e-prints}, January 2018.

\bibitem[{Ghisellini} and {Tavecchio}(2009)]{GGTF09}
G.~{Ghisellini} and F.~{Tavecchio}.
\newblock {Canonical high-power blazars}.
\newblock \emph{Monthly Notices of the RAS}, 397:\penalty0 985--1002, August
  2009.
\newblock \doi{10.1111/j.1365-2966.2009.15007.x}.

\bibitem[{Ghisellini} and {Tavecchio}(2015)]{GGFT15}
G.~{Ghisellini} and F.~{Tavecchio}.
\newblock {Fermi/LAT broad emission line blazars}.
\newblock \emph{\mnras}, 448\penalty0 (2):\penalty0 1060--1077, April 2015.
\newblock \doi{10.1093/mnras/stv055}.

\bibitem[{Ghisellini} et~al.(2014){Ghisellini}, {Tavecchio}, {Maraschi},
  {Celotti}, and {Sbarrato}]{GGNature}
G.~{Ghisellini}, F.~{Tavecchio}, L.~{Maraschi}, A.~{Celotti}, and
  T.~{Sbarrato}.
\newblock {The power of relativistic jets is larger than the luminosity of
  their accretion disks}.
\newblock \emph{\nat}, 515\penalty0 (7527):\penalty0 376--378, November 2014.
\newblock \doi{10.1038/nature13856}.

\bibitem[{Ghisellini} et~al.(2017){Ghisellini}, {Righi}, {Costamante}, and
  {Tavecchio}]{GG17}
G.~{Ghisellini}, C.~{Righi}, L.~{Costamante}, and F.~{Tavecchio}.
\newblock {The Fermi blazar sequence}.
\newblock \emph{Monthly Notices of the RAS}, 469:\penalty0 255--266, July 2017.
\newblock \doi{10.1093/mnras/stx806}.

\bibitem[{H.~E.~S.~S. Collaboration} et~al.(2019){H.~E.~S.~S. Collaboration},
  {Abdalla}, {Adam}, {Aharonian}, {Ait Benkhali}, {Ang{\"u}ner}, {Arakawa},
  {Arcaro}, {Armand}, {Ashkar}, {Backes}, {Barbosa Martins}, {Barnard},
  {Becherini}, {Berge}, {Bernl{\"o}hr}, {Blackwell}, {B{\"o}ttcher}, {Boisson},
  {Bolmont}, {Bonnefoy}, {Bregeon}, {Breuhaus}, {Brun}, {Brun}, {Bryan},
  {B{\"u}chele}, {Bulik}, {Bylund}, {Capasso}, {Caroff}, {Carosi}, {Casanova},
  {Cerruti}, {Chand}, {Chandra}, {Chen}, {Colafrancesco}, {Cury{\l}o},
  {Davids}, {Deil}, {Devin}, {deWilt}, {Dirson}, {Djannati-Ata{\"\i}},
  {Dmytriiev}, {Donath}, {Doroshenko}, {Drury}, {Dyks}, {Egberts}, {Emery},
  {Ernenwein}, {Eschbach}, {Feijen}, {Fegan}, {Fiasson}, {Fontaine}, {Funk},
  {F{\"u}{\ss}ling}, {Gabici}, {Gallant}, {Gat{\'e}}, {Giavitto}, {Glawion},
  {Glicenstein}, {Gottschall}, {Grondin}, {Hahn}, {Haupt}, {Heinzelmann},
  {Henri}, {Hermann}, {Hinton}, {Hofmann}, {Hoischen}, {Holch}, {Holler},
  {Horns}, {Huber}, {Iwasaki}, {Jamrozy}, {Jankowsky}, {Jankowsky},
  {Jardin-Blicq}, {Jung-Richardt}, {Kastendieck}, {Katarzy{\'n}ski},
  {Katsuragawa}, {Katz}, {Khangulyan}, {Kh{\'e}lifi}, {King}, {Klepser},
  {Klu{\'z}niak}, {Komin}, {Kosack}, {Kostunin}, {Kraus}, {Lamanna}, {Lau},
  {Lemi{\`e}re}, {Lemoine-Goumard}, {Lenain}, {Leser}, {Levy}, {Lohse},
  {Lypova}, {Mackey}, {Majumdar}, {Malyshev}, {Marandon}, {Marcowith}, {Mares},
  {Mariaud}, {Mart{\'\i}-Devesa}, {Marx}, {Maurin}, {Meintjes}, {Mitchell},
  {Moderski}, {Mohamed}, {Mohrmann}, {Moore}, {Moulin}, {Muller}, {Murach},
  {Nakashima}, {de Naurois}, {Ndiyavala}, {Niederwanger}, {Niemiec}, {Oakes},
  {O'Brien}, {Odaka}, {Ohm}, {de Ona Wilhelmi}, {Ostrowski}, {Oya}, {Panter},
  {Parsons}, {Perennes}, {Petrucci}, {Peyaud}, {Piel}, {Pita}, {Poireau},
  {Priyana Noel}, {Prokhorov}, {Prokoph}, {P{\"u}hlhofer}, {Punch},
  {Quirrenbach}, {Raab}, {Rauth}, {Reimer}, {Reimer}, {Remy}, {Renaud},
  {Rieger}, {Rinchiuso}, {Romoli}, {Rowell}, {Rudak}, {Ruiz-Velasco},
  {Sahakian}, {Saito}, {Sanchez}, {Santangelo}, {Sasaki}, {Schlickeiser},
  {Sch{\"u}ssler}, {Schulz}, {Schutte}, {Schwanke}, {Schwemmer},
  {Seglar-Arroyo}, {Senniappan}, {Seyffert}, {Shafi}, {Shiningayamwe},
  {Simoni}, {Sinha}, {Sol}, {Specovius}, {Spir-Jacob}, {Stawarz}, {Steenkamp},
  {Stegmann}, {Steppa}, {Takahashi}, {Tavernier}, {Taylor}, {Terrier},
  {Tiziani}, {Tluczykont}, {Trichard}, {Tsirou}, {Tsuji}, {Tuffs}, {Uchiyama},
  {van der Walt}, {van Eldik}, {van Rensburg}, {van Soelen}, {Vasileiadis},
  {Veh}, {Venter}, {Vincent}, {Vink}, {Voisin}, {V{\"o}lk}, {Vuillaume},
  {Wadiasingh}, {Wagner}, {White}, {Wierzcholska}, {Yang}, {Yoneda},
  {Zacharias}, {Zanin}, {Zdziarski}, {Zech}, {Ziegler}, {Zorn}, {{\.Z}ywucka},
  and {Meyer}]{HESS_3C279}
{H.~E.~S.~S. Collaboration}, H.~{Abdalla}, R.~{Adam}, F.~{Aharonian}, F.~{Ait
  Benkhali}, E.~O. {Ang{\"u}ner}, M.~{Arakawa}, C.~{Arcaro}, C.~{Armand},
  H.~{Ashkar}, M.~{Backes}, V.~{Barbosa Martins}, M.~{Barnard}, Y.~{Becherini},
  D.~{Berge}, K.~{Bernl{\"o}hr}, R.~{Blackwell}, M.~{B{\"o}ttcher},
  C.~{Boisson}, J.~{Bolmont}, S.~{Bonnefoy}, J.~{Bregeon}, M.~{Breuhaus},
  F.~{Brun}, P.~{Brun}, M.~{Bryan}, M.~{B{\"u}chele}, T.~{Bulik}, T.~{Bylund},
  M.~{Capasso}, S.~{Caroff}, A.~{Carosi}, S.~{Casanova}, M.~{Cerruti},
  T.~{Chand}, S.~{Chandra}, A.~{Chen}, S.~{Colafrancesco}, M.~{Cury{\l}o},
  I.~D. {Davids}, C.~{Deil}, J.~{Devin}, P.~{deWilt}, L.~{Dirson},
  A.~{Djannati-Ata{\"\i}}, A.~{Dmytriiev}, A.~{Donath}, V.~{Doroshenko},
  L.~O.~'C. {Drury}, J.~{Dyks}, K.~{Egberts}, G.~{Emery}, J.~P. {Ernenwein},
  S.~{Eschbach}, K.~{Feijen}, S.~{Fegan}, A.~{Fiasson}, G.~{Fontaine},
  S.~{Funk}, M.~{F{\"u}{\ss}ling}, S.~{Gabici}, Y.~A. {Gallant}, F.~{Gat{\'e}},
  G.~{Giavitto}, D.~{Glawion}, J.~F. {Glicenstein}, D.~{Gottschall}, M.~H.
  {Grondin}, J.~{Hahn}, M.~{Haupt}, G.~{Heinzelmann}, G.~{Henri}, G.~{Hermann},
  J.~A. {Hinton}, W.~{Hofmann}, C.~{Hoischen}, T.~L. {Holch}, M.~{Holler},
  D.~{Horns}, D.~{Huber}, H.~{Iwasaki}, M.~{Jamrozy}, D.~{Jankowsky},
  F.~{Jankowsky}, A.~{Jardin-Blicq}, I.~{Jung-Richardt}, M.~A. {Kastendieck},
  K.~{Katarzy{\'n}ski}, M.~{Katsuragawa}, U.~{Katz}, D.~{Khangulyan},
  B.~{Kh{\'e}lifi}, J.~{King}, S.~{Klepser}, W.~{Klu{\'z}niak}, Nu. {Komin},
  K.~{Kosack}, D.~{Kostunin}, M.~{Kraus}, G.~{Lamanna}, J.~{Lau},
  A.~{Lemi{\`e}re}, M.~{Lemoine-Goumard}, J.~P. {Lenain}, E.~{Leser},
  C.~{Levy}, T.~{Lohse}, I.~{Lypova}, J.~{Mackey}, J.~{Majumdar},
  D.~{Malyshev}, V.~{Marandon}, A.~{Marcowith}, A.~{Mares}, C.~{Mariaud},
  G.~{Mart{\'\i}-Devesa}, R.~{Marx}, G.~{Maurin}, P.~J. {Meintjes}, A.~M.~W.
  {Mitchell}, R.~{Moderski}, M.~{Mohamed}, L.~{Mohrmann}, C.~{Moore},
  E.~{Moulin}, J.~{Muller}, T.~{Murach}, S.~{Nakashima}, M.~{de Naurois},
  H.~{Ndiyavala}, F.~{Niederwanger}, J.~{Niemiec}, L.~{Oakes}, P.~{O'Brien},
  H.~{Odaka}, S.~{Ohm}, E.~{de Ona Wilhelmi}, M.~{Ostrowski}, I.~{Oya},
  M.~{Panter}, R.~D. {Parsons}, C.~{Perennes}, P.~O. {Petrucci}, B.~{Peyaud},
  Q.~{Piel}, S.~{Pita}, V.~{Poireau}, A.~{Priyana Noel}, D.~A. {Prokhorov},
  H.~{Prokoph}, G.~{P{\"u}hlhofer}, M.~{Punch}, A.~{Quirrenbach}, S.~{Raab},
  R.~{Rauth}, A.~{Reimer}, O.~{Reimer}, Q.~{Remy}, M.~{Renaud}, F.~{Rieger},
  L.~{Rinchiuso}, C.~{Romoli}, G.~{Rowell}, B.~{Rudak}, E.~{Ruiz-Velasco},
  V.~{Sahakian}, S.~{Saito}, D.~A. {Sanchez}, A.~{Santangelo}, M.~{Sasaki},
  R.~{Schlickeiser}, F.~{Sch{\"u}ssler}, A.~{Schulz}, H.~{Schutte},
  U.~{Schwanke}, S.~{Schwemmer}, M.~{Seglar-Arroyo}, M.~{Senniappan}, A.~S.
  {Seyffert}, N.~{Shafi}, K.~{Shiningayamwe}, R.~{Simoni}, A.~{Sinha},
  H.~{Sol}, A.~{Specovius}, M.~{Spir-Jacob}, L.~{Stawarz}, R.~{Steenkamp},
  C.~{Stegmann}, C.~{Steppa}, T.~{Takahashi}, T.~{Tavernier}, A.~M. {Taylor},
  R.~{Terrier}, D.~{Tiziani}, M.~{Tluczykont}, C.~{Trichard}, M.~{Tsirou},
  N.~{Tsuji}, R.~{Tuffs}, Y.~{Uchiyama}, D.~J. {van der Walt}, C.~{van Eldik},
  C.~{van Rensburg}, B.~{van Soelen}, G.~{Vasileiadis}, J.~{Veh}, C.~{Venter},
  P.~{Vincent}, J.~{Vink}, F.~{Voisin}, H.~J. {V{\"o}lk}, T.~{Vuillaume},
  Z.~{Wadiasingh}, S.~J. {Wagner}, R.~{White}, A.~{Wierzcholska}, R.~{Yang},
  H.~{Yoneda}, M.~{Zacharias}, R.~{Zanin}, A.~A. {Zdziarski}, A.~{Zech},
  A.~{Ziegler}, J.~{Zorn}, N.~{{\.Z}ywucka}, and M.~{Meyer}.
\newblock {Constraints on the emission region of <ASTROBJ>3C 279</ASTROBJ>
  during strong flares in 2014 and 2015 through VHE {\ensuremath{\gamma}}-ray
  observations with H.E.S.S.}
\newblock \emph{\aap}, 627:\penalty0 A159, July 2019.
\newblock \doi{10.1051/0004-6361/201935704}.

\bibitem[Kadler et~al.(2016)]{Kadler:2016ygj}
M.~Kadler et~al.
\newblock {Coincidence of a high-fluence blazar outburst with a PeV-energy
  neutrino event}.
\newblock \emph{Nature Phys.}, 12\penalty0 (8):\penalty0 807--814, 2016.
\newblock \doi{10.1038/nphys3715, 10.1038/NPHYS3715}.

\bibitem[{Kaspi} et~al.(2007){Kaspi}, {Brandt}, {Maoz}, {Netzer}, {Schneider},
  and {Shemmer}]{Kaspi07}
Shai {Kaspi}, W.~N. {Brandt}, Dan {Maoz}, Hagai {Netzer}, Donald~P.
  {Schneider}, and Ohad {Shemmer}.
\newblock {Reverberation Mapping of High-Luminosity Quasars: First Results}.
\newblock \emph{\apj}, 659\penalty0 (2):\penalty0 997--1007, April 2007.
\newblock \doi{10.1086/512094}.

\bibitem[{Komissarov} et~al.(2007){Komissarov}, {Barkov}, {Vlahakis}, and
  {K{\"o}nigl}]{Komissarov07}
Serguei~S. {Komissarov}, Maxim~V. {Barkov}, Nektarios {Vlahakis}, and Arieh
  {K{\"o}nigl}.
\newblock {Magnetic acceleration of relativistic active galactic nucleus jets}.
\newblock \emph{\mnras}, 380\penalty0 (1):\penalty0 51--70, September 2007.
\newblock \doi{10.1111/j.1365-2966.2007.12050.x}.

\bibitem[{MAGIC Collaboration} et~al.(2018){MAGIC Collaboration}, {Acciari},
  {Ansoldi}, {Antonelli}, {Arbet Engels}, {Arcaro}, {Baack}, {Babi{\'c}},
  {Banerjee}, {Bangale}, {Barres de Almeida}, {Barrio}, {Bednarek},
  {Bernardini}, {Berti}, {Besenrieder}, {Bhattacharyya}, {Bigongiari},
  {Biland}, {Blanch}, {Bonnoli}, {Carosi}, {Ceribella}, {Cikota}, {Colak},
  {Colin}, {Colombo}, {Contreras}, {Cortina}, {Covino}, {D'Elia}, {da Vela},
  {Dazzi}, {de Angelis}, {de Lotto}, {Delfino}, {Delgado}, {di Pierro}, {Do
  Souto Espi{\~n}era}, {Dom{\'\i}nguez}, {Dominis Prester}, {Dorner}, {Doro},
  {Einecke}, {Elsaesser}, {Fallah Ramazani}, {Fattorini},
  {Fern{\'a}ndez-Barral}, {Ferrara}, {Fidalgo}, {Foffano}, {Fonseca}, {Font},
  {Fruck}, {Galindo}, {Gallozzi}, {Garc{\'\i}a L{\'o}pez}, {Garczarczyk},
  {Gaug}, {Giammaria}, {Godinovi{\'c}}, {Guberman}, {Hadasch}, {Hahn},
  {Hassan}, {Herrera}, {Hoang}, {Hrupec}, {Inoue}, {Ishio}, {Iwamura}, {Kubo},
  {Kushida}, {Kuve{\v{z}}di{\'c}}, {Lamastra}, {Lelas}, {Leone}, {Lindfors},
  {Lombardi}, {Longo}, {L{\'o}pez}, {L{\'o}pez-Oramas}, {Maggio}, {Majumdar},
  {Makariev}, {Maneva}, {Manganaro}, {Mannheim}, {Maraschi}, {Mariotti},
  {Mart{\'\i}nez}, {Masuda}, {Mazin}, {Minev}, {Miranda}, {Mirzoyan}, {Molina},
  {Moralejo}, {Moreno}, {Moretti}, {Munar-Adrover}, {Neustroev}, {Niedzwiecki},
  {Nievas Rosillo}, {Nigro}, {Nilsson}, {Ninci}, {Nishijima}, {Noda},
  {Nogu{\'e}s}, {Paiano}, {Palacio}, {Paneque}, {Paoletti}, {Paredes},
  {Pedaletti}, {Pe{\~n}il}, {Peresano}, {Persic}, {Prada Moroni}, {Prand ini},
  {Puljak}, {Garcia}, {Rhode}, {Rib{\'o}}, {Rico}, {Righi}, {Rugliancich},
  {Saha}, {Saito}, {Satalecka}, {Schweizer}, {Sitarek}, {{\v{S}}nidari{\'c}},
  {Sobczynska}, {Somero}, {Stamerra}, {Strzys}, {Suri{\'c}}, {Tavecchio},
  {Temnikov}, {Terzi{\'c}}, {Teshima}, {Torres-Alb{\`a}}, {Tsujimoto}, {van
  Scherpenberg}, {Vanzo}, {Vazquez Acosta}, {Vovk}, {Ward}, {Will},
  {Zari{\'c}}, {Fermi-Lat Collaboration}, {Becerra Gonz{\'a}lez}, {Raiteri},
  {Sandrinelli}, {Hovatta}, {Kiehlmann}, {Max-Moerbeck}, {Tornikoski},
  {L{\"a}hteenm{\"a}ki}, {Tammi}, {Ramakrishnan}, {Thum}, {Agudo}, {Molina},
  {G{\'o}mez}, {Fuentes}, {Casadio}, {Traianou}, {Myserlis}, and
  {Kim}]{MAGIC_PKS1510-089}
{MAGIC Collaboration}, V.~A. {Acciari}, S.~{Ansoldi}, L.~A. {Antonelli},
  A.~{Arbet Engels}, C.~{Arcaro}, D.~{Baack}, A.~{Babi{\'c}}, B.~{Banerjee},
  P.~{Bangale}, U.~{Barres de Almeida}, J.~A. {Barrio}, W.~{Bednarek},
  E.~{Bernardini}, A.~{Berti}, J.~{Besenrieder}, W.~{Bhattacharyya},
  C.~{Bigongiari}, A.~{Biland}, O.~{Blanch}, G.~{Bonnoli}, R.~{Carosi},
  G.~{Ceribella}, S.~{Cikota}, S.~M. {Colak}, P.~{Colin}, E.~{Colombo}, J.~L.
  {Contreras}, J.~{Cortina}, S.~{Covino}, V.~{D'Elia}, P.~{da Vela},
  F.~{Dazzi}, A.~{de Angelis}, B.~{de Lotto}, M.~{Delfino}, J.~{Delgado},
  F.~{di Pierro}, E.~{Do Souto Espi{\~n}era}, A.~{Dom{\'\i}nguez}, D.~{Dominis
  Prester}, D.~{Dorner}, M.~{Doro}, S.~{Einecke}, D.~{Elsaesser}, V.~{Fallah
  Ramazani}, A.~{Fattorini}, A.~{Fern{\'a}ndez-Barral}, G.~{Ferrara},
  D.~{Fidalgo}, L.~{Foffano}, M.~V. {Fonseca}, L.~{Font}, C.~{Fruck},
  D.~{Galindo}, S.~{Gallozzi}, R.~J. {Garc{\'\i}a L{\'o}pez}, M.~{Garczarczyk},
  M.~{Gaug}, P.~{Giammaria}, N.~{Godinovi{\'c}}, D.~{Guberman}, D.~{Hadasch},
  A.~{Hahn}, T.~{Hassan}, J.~{Herrera}, J.~{Hoang}, D.~{Hrupec}, S.~{Inoue},
  K.~{Ishio}, Y.~{Iwamura}, H.~{Kubo}, J.~{Kushida}, D.~{Kuve{\v{z}}di{\'c}},
  A.~{Lamastra}, D.~{Lelas}, F.~{Leone}, E.~{Lindfors}, S.~{Lombardi},
  F.~{Longo}, M.~{L{\'o}pez}, A.~{L{\'o}pez-Oramas}, C.~{Maggio},
  P.~{Majumdar}, M.~{Makariev}, G.~{Maneva}, M.~{Manganaro}, K.~{Mannheim},
  L.~{Maraschi}, M.~{Mariotti}, M.~{Mart{\'\i}nez}, S.~{Masuda}, D.~{Mazin},
  M.~{Minev}, J.~M. {Miranda}, R.~{Mirzoyan}, E.~{Molina}, A.~{Moralejo},
  V.~{Moreno}, E.~{Moretti}, P.~{Munar-Adrover}, V.~{Neustroev},
  A.~{Niedzwiecki}, M.~{Nievas Rosillo}, C.~{Nigro}, K.~{Nilsson}, D.~{Ninci},
  K.~{Nishijima}, K.~{Noda}, L.~{Nogu{\'e}s}, S.~{Paiano}, J.~{Palacio},
  D.~{Paneque}, R.~{Paoletti}, J.~M. {Paredes}, G.~{Pedaletti}, P.~{Pe{\~n}il},
  M.~{Peresano}, M.~{Persic}, P.~G. {Prada Moroni}, E.~{Prand ini},
  I.~{Puljak}, J.~R. {Garcia}, W.~{Rhode}, M.~{Rib{\'o}}, J.~{Rico},
  C.~{Righi}, A.~{Rugliancich}, L.~{Saha}, T.~{Saito}, K.~{Satalecka},
  T.~{Schweizer}, J.~{Sitarek}, I.~{{\v{S}}nidari{\'c}}, D.~{Sobczynska},
  A.~{Somero}, A.~{Stamerra}, M.~{Strzys}, T.~{Suri{\'c}}, F.~{Tavecchio},
  P.~{Temnikov}, T.~{Terzi{\'c}}, M.~{Teshima}, N.~{Torres-Alb{\`a}},
  S.~{Tsujimoto}, J.~{van Scherpenberg}, G.~{Vanzo}, M.~{Vazquez Acosta},
  I.~{Vovk}, J.~E. {Ward}, M.~{Will}, D.~{Zari{\'c}}, {Fermi-Lat
  Collaboration}, J.~{Becerra Gonz{\'a}lez}, C.~M. {Raiteri}, A.~{Sandrinelli},
  T.~{Hovatta}, S.~{Kiehlmann}, W.~{Max-Moerbeck}, M.~{Tornikoski},
  A.~{L{\"a}hteenm{\"a}ki}, J.~{Tammi}, V.~{Ramakrishnan}, C.~{Thum},
  I.~{Agudo}, S.~N. {Molina}, J.~L. {G{\'o}mez}, A.~{Fuentes}, C.~{Casadio},
  E.~{Traianou}, I.~{Myserlis}, and J.~Y. {Kim}.
\newblock {Detection of persistent VHE gamma-ray emission from PKS 1510-089 by
  the MAGIC telescopes during low states between 2012 and 2017}.
\newblock \emph{\aap}, 619:\penalty0 A159, November 2018.
\newblock \doi{10.1051/0004-6361/201833618}.

\bibitem[{McKinney}(2006)]{McKinney06}
Jonathan~C. {McKinney}.
\newblock {General relativistic magnetohydrodynamic simulations of the jet
  formation and large-scale propagation from black hole accretion systems}.
\newblock \emph{\mnras}, 368\penalty0 (4):\penalty0 1561--1582, June 2006.
\newblock \doi{10.1111/j.1365-2966.2006.10256.x}.

\bibitem[{M{\'e}sz{\'a}ros}(2017)]{Meszaros17}
P.~{M{\'e}sz{\'a}ros}.
\newblock {Astrophysical Sources of High-Energy Neutrinos in the IceCube Era}.
\newblock \emph{Annual Review of Nuclear and Particle Science}, 67:\penalty0
  45--67, October 2017.
\newblock \doi{10.1146/annurev-nucl-101916-123304}.

\bibitem[{M{\"u}cke} et~al.(2003){M{\"u}cke}, {Protheroe}, {Engel}, {Rachen},
  and {Stanev}]{Mucke2003}
A.~{M{\"u}cke}, R.~J. {Protheroe}, R.~{Engel}, J.~P. {Rachen}, and T.~{Stanev}.
\newblock {BL Lac objects in the synchrotron proton blazar model}.
\newblock \emph{Astroparticle Physics}, 18\penalty0 (6):\penalty0 593--613,
  March 2003.
\newblock \doi{10.1016/S0927-6505(02)00185-8}.

\bibitem[{Murase} and {Waxman}(2016)]{MuraseWaxman16}
K.~{Murase} and E.~{Waxman}.
\newblock {Constraining high-energy cosmic neutrino sources: Implications and
  prospects}.
\newblock \emph{Physical Review D}, 94\penalty0 (10):\penalty0 103006, November
  2016.
\newblock \doi{10.1103/PhysRevD.94.103006}.

\bibitem[{Murase} et~al.(2014){Murase}, {Inoue}, and {Dermer}]{Murase14}
K.~{Murase}, Y.~{Inoue}, and C.~D. {Dermer}.
\newblock {Diffuse neutrino intensity from the inner jets of active galactic
  nuclei: Impacts of external photon fields and the blazar sequence}.
\newblock \emph{Physical Review D}, 90\penalty0 (2):\penalty0 023007, July
  2014.
\newblock \doi{10.1103/PhysRevD.90.023007}.

\bibitem[{Murase} et~al.(2012){Murase}, {Beacom}, and {Takami}]{Murase12}
Kohta {Murase}, John~F. {Beacom}, and Hajime {Takami}.
\newblock {Gamma-ray and neutrino backgrounds as probes of the high-energy
  universe: hints of cascades, general constraints, and implications for TeV
  searches}.
\newblock \emph{\jcap}, 2012\penalty0 (8):\penalty0 030, August 2012.
\newblock \doi{10.1088/1475-7516/2012/08/030}.

\bibitem[Murase et~al.(2018)Murase, Oikonomou, and Petropoulou]{Murase:2018iyl}
Kohta Murase, Foteini Oikonomou, and Maria Petropoulou.
\newblock {Blazar Flares as an Origin of High-Energy Cosmic Neutrinos?}
\newblock \emph{Astrophys. J.}, 865\penalty0 (2):\penalty0 124, 2018.
\newblock \doi{10.3847/1538-4357/aada00}.

\bibitem[{Nakamura} et~al.(2018){Nakamura}, {Asada}, {Hada}, {Pu}, {Noble},
  {Tseng}, {Toma}, {Kino}, {Nagai}, {Takahashi}, {Algaba}, {Orienti},
  {Akiyama}, {Doi}, {Giovannini}, {Giroletti}, {Honma}, {Koyama}, {Lico},
  {Niinuma}, and {Tazaki}]{Nakamura}
Masanori {Nakamura}, Keiichi {Asada}, Kazuhiro {Hada}, Hung-Yi {Pu}, Scott
  {Noble}, Chihyin {Tseng}, Kenji {Toma}, Motoki {Kino}, Hiroshi {Nagai},
  Kazuya {Takahashi}, Juan-Carlos {Algaba}, Monica {Orienti}, Kazunori
  {Akiyama}, Akihiro {Doi}, Gabriele {Giovannini}, Marcello {Giroletti}, Mareki
  {Honma}, Shoko {Koyama}, Rocco {Lico}, Kotaro {Niinuma}, and Fumie {Tazaki}.
\newblock {Parabolic Jets from the Spinning Black Hole in M87}.
\newblock \emph{\apj}, 868\penalty0 (2):\penalty0 146, December 2018.
\newblock \doi{10.3847/1538-4357/aaeb2d}.

\bibitem[Padovani et~al.(2016)Padovani, Resconi, Giommi, Arsioli, and
  Chang]{Padovani:2016wwn}
P.~Padovani, E.~Resconi, P.~Giommi, B.~Arsioli, and Y.~L. Chang.
\newblock {Extreme blazars as counterparts of IceCube astrophysical neutrinos}.
\newblock \emph{Mon. Not. Roy. Astron. Soc.}, 457\penalty0 (4):\penalty0
  3582--3592, 2016.
\newblock \doi{10.1093/mnras/stw228}.

\bibitem[Padovani et~al.(2019)Padovani, Oikonomou, Petropoulou, Giommi, and
  Resconi]{Padovani:2019xcv}
P.~Padovani, F.~Oikonomou, M.~Petropoulou, P.~Giommi, and E.~Resconi.
\newblock {TXS 0506+056, the first cosmic neutrino source, is not a BL Lac}.
\newblock \emph{Mon. Not. Roy. Astron. Soc.}, 484\penalty0 (1):\penalty0
  L104--L108, 2019.
\newblock \doi{10.1093/mnrasl/slz011}.

\bibitem[Palladino et~al.(2019)Palladino, Rodrigues, Gao, and
  Winter]{Palladino:2018lov}
Andrea Palladino, Xavier Rodrigues, Shan Gao, and Walter Winter.
\newblock {Interpretation of the diffuse astrophysical neutrino flux in terms
  of the blazar sequence}.
\newblock \emph{Astrophys. J.}, 871\penalty0 (1):\penalty0 41, 2019.
\newblock \doi{10.3847/1538-4357/aaf507}.

\bibitem[{Poutanen} and {Stern}(2010)]{Poutanen10}
Juri {Poutanen} and Boris {Stern}.
\newblock {GeV Breaks in Blazars as a Result of Gamma-ray Absorption Within the
  Broad-line Region}.
\newblock \emph{\apjl}, 717\penalty0 (2):\penalty0 L118--L121, July 2010.
\newblock \doi{10.1088/2041-8205/717/2/L118}.

\bibitem[{Rieger} et~al.(2007){Rieger}, {Bosch-Ramon}, and {Duffy}]{Rieger07}
Frank~M. {Rieger}, Valent{\'\i} {Bosch-Ramon}, and Peter {Duffy}.
\newblock {Fermi acceleration in astrophysical jets}.
\newblock \emph{\apss}, 309\penalty0 (1-4):\penalty0 119--125, June 2007.
\newblock \doi{10.1007/s10509-007-9466-z}.

\bibitem[Righi et~al.(2017)Righi, Tavecchio, and Guetta]{Righi:2016kio}
Chiara Righi, Fabrizio Tavecchio, and Dafne Guetta.
\newblock {High-energy emitting BL Lacs and high-energy neutrinos - Prospects
  for the direct association with IceCube and KM3NeT}.
\newblock \emph{Astron. Astrophys.}, 598:\penalty0 A36, 2017.
\newblock \doi{10.1051/0004-6361/201629412}.

\bibitem[{Rodrigues} et~al.(2018){Rodrigues}, {Fedynitch}, {Gao}, {Boncioli},
  and {Winter}]{Rodrigues2017}
Xavier {Rodrigues}, Anatoli {Fedynitch}, Shan {Gao}, Denise {Boncioli}, and
  Walter {Winter}.
\newblock {Neutrinos and Ultra-high-energy Cosmic-ray Nuclei from Blazars}.
\newblock \emph{\apj}, 854\penalty0 (1):\penalty0 54, February 2018.
\newblock \doi{10.3847/1538-4357/aaa7ee}.

\bibitem[{Romero} et~al.(2017{\natexlab{a}}){Romero}, {Boettcher}, {Markoff},
  and {Tavecchio}]{Romero}
Gustavo~E. {Romero}, M.~{Boettcher}, S.~{Markoff}, and F.~{Tavecchio}.
\newblock {Relativistic Jets in Active Galactic Nuclei and Microquasars}.
\newblock \emph{\ssr}, 207\penalty0 (1-4):\penalty0 5--61, July
  2017{\natexlab{a}}.
\newblock \doi{10.1007/s11214-016-0328-2}.

\bibitem[{Romero} et~al.(2017{\natexlab{b}}){Romero}, {Boettcher}, {Markoff},
  and {Tavecchio}]{Romero17}
Gustavo~E. {Romero}, M.~{Boettcher}, S.~{Markoff}, and F.~{Tavecchio}.
\newblock {Relativistic Jets in Active Galactic Nuclei and Microquasars}.
\newblock \emph{\ssr}, 207\penalty0 (1-4):\penalty0 5--61, July
  2017{\natexlab{b}}.
\newblock \doi{10.1007/s11214-016-0328-2}.

\bibitem[{Sahakyan}(2018)]{Sahakyan18}
N.~{Sahakyan}.
\newblock {Lepto-hadronic $\gamma$-ray and neutrino emission from the jet of
  TXS 0506+056}.
\newblock \emph{ArXiv e-prints}, August 2018.

\bibitem[{Shakura} and {Sunyaev}(1973)]{Shakura97}
N.~I. {Shakura} and R.~A. {Sunyaev}.
\newblock {Black holes in binary systems. Observational appearance.}
\newblock \emph{Astronomy and Astrophysics}, 24:\penalty0 337--355, 1973.

\bibitem[{Sikora} et~al.(1994){Sikora}, {Begelman}, and {Rees}]{Sikora94}
M.~{Sikora}, M.~C. {Begelman}, and M.~J. {Rees}.
\newblock {Comptonization of diffuse ambient radiation by a relativistic jet:
  The source of gamma rays from blazars?}
\newblock \emph{Astrophysical Journal}, 421:\penalty0 153--162, January 1994.
\newblock \doi{10.1086/173633}.

\bibitem[{Tavecchio} and {Ghisellini}(2015)]{Tavecchio15}
F.~{Tavecchio} and G.~{Ghisellini}.
\newblock {High-energy cosmic neutrinos from spine-sheath BL Lac jets}.
\newblock \emph{Monthly Notices of the RAS}, 451:\penalty0 1502--1510, August
  2015.
\newblock \doi{10.1093/mnras/stv1023}.

\bibitem[{Tavecchio} et~al.(2011){Tavecchio}, {Becerra-Gonzalez}, {Ghisellini},
  {Stamerra}, {Bonnoli}, {Foschini}, and {Maraschi}]{FT11}
F.~{Tavecchio}, J.~{Becerra-Gonzalez}, G.~{Ghisellini}, A.~{Stamerra},
  G.~{Bonnoli}, L.~{Foschini}, and L.~{Maraschi}.
\newblock {On the origin of the {\ensuremath{\gamma}}-ray emission from the
  flaring blazar PKS 1222+216}.
\newblock \emph{\aap}, 534:\penalty0 A86, October 2011.
\newblock \doi{10.1051/0004-6361/201117204}.

\bibitem[{Tavecchio} et~al.(2014){Tavecchio}, {Ghisellini}, and
  {Guetta}]{Tavecchio14}
F.~{Tavecchio}, G.~{Ghisellini}, and D.~{Guetta}.
\newblock {Structured Jets in BL Lac Objects: Efficient PeV Neutrino
  Factories?}
\newblock \emph{Astrophysical Journal, Letters}, 793:\penalty0 L18, September
  2014.
\newblock \doi{10.1088/2041-8205/793/1/L18}.

\bibitem[{Tavecchio} and {Ghisellini}(2008)]{FTGG08}
Fabrizio {Tavecchio} and Gabriele {Ghisellini}.
\newblock {The spectrum of the broad-line region and the high-energy emission
  of powerful blazars}.
\newblock \emph{\mnras}, 386\penalty0 (2):\penalty0 945--952, May 2008.
\newblock \doi{10.1111/j.1365-2966.2008.13072.x}.

\bibitem[{Vlahakis}(2015)]{Vlahakis15}
Nektarios {Vlahakis}.
\newblock \emph{{Theory of Relativistic Jets}}, volume 414 of
  \emph{Astrophysics and Space Science Library}, page 177.
\newblock 2015.
\newblock \doi{10.1007/978-3-319-10356-3_7}.

\end{thebibliography}
\bibliographystyle{plainnat}

\end{document}